\documentclass[onecolumn,conference,moreauthors, a4paper]{IEEEtran}
\usepackage{filecontents}
\usepackage{cite}
\usepackage{amssymb,amsmath,epsfig,latexsym,graphicx,bm}
\usepackage{tabu}
\usepackage{tabularx}
\usepackage{graphicx,bm,makecell,array,optidef}
\usepackage[justification=centering]{caption}
\usepackage{subcaption}
\usepackage{float}
\usepackage{tabularx}
\usepackage{amsmath,amssymb,amsfonts}
\usepackage{algorithmic}
\usepackage{amsmath}
\usepackage[left = 1 in, top=1 in,right=1 in,bottom=1 in,nohead,nofoot]{geometry}
\usepackage{textcomp}
\usepackage{hyperref}
\usepackage{multirow}
\usepackage{array}
\usepackage{makecell}
\usepackage{ragged2e}
\usepackage{epstopdf}
\usepackage{array}
\usepackage{longtable}
\usepackage{lipsum}
\usepackage{ragged2e}
\usepackage[T1]{fontenc}
\usepackage{lmodern}

\title{The Convergence of Blockchain, IoT and 6G: Potential, Opportunities, Challenges and Research Roadmap}
\newcommand{\orcid}[1]{\href{https://orcid.org/#1}{\textcolor[HTML]{A6CE39}}}

\author{\IEEEauthorblockN{Abu Jahid, Mohammed H. Alsharif and Trevor J. Hall \\
\centering }}
\providecommand{\keywords}[1]{\textbf{\textit{Keywords---}} #1}
\begin{document}
\bstctlcite{IEEEexample:BSTcontrol}
\maketitle

\begin{abstract}

The world is undergoing a profound transformation with the advent of intelligent information era. The fundamental domains linked with smart living such as transportation, entertainment, healthcare, and smart cities are anticipated to elevate the level of service quality assuring a high-end user experience. 6G networks envisioned being the game changer in next generation wireless communication systems that will address the challenges of limited information speed escalated with the augmentation of billions of data applications encountered by the current fifth generation (5G) networks. Some key radical technologies in 6G together with existing 5G candidate schemes will guarantee the expected quality of experience (QoE) to attain ubiquitous wireless connectivity for the Internet of Everything (IoE) ranging from the telecom industry to digital smart industries. Recently, Internet of Things (IoT) is reforming the incumbent industry into smart infrastructure featured with 6G data-driven architecture. Blockchain technology (BCT) has gained significant attention due to undertake the decentralization, transparency, spectrum resource scarcity, inherent privacy and security, poor interoperability, confidentiality, and emerging smart applications domains including Industrial IoT and Industry 4.0. The mismatch between the requirements of many data intensive disruptive IoT applications and 5G network capabilities steered the demand of decentralized BCT based 6G architecture. Inspired by these facts, this paper studies an extensive survey to draw a new direction of blockchain integration into 6G mobile networks, IoT technologies, and smart industries focusing the potential merits and challenges in terms of infrastructure sharing, computational loads, latency, bandwidth overhead, business model, sustainability goals, and edge intelligence. We highlighted the convergence of IoT in blockchain to enable intelligent distribution in future industrial IoT and the technical model of 6G networks to realize the successful deployment of BCT schemes. This paper pointed out the current intriguing challenges, canvassed the mitigation techniques, and plausible future research opportunities that may benefit the pursuit of this vision.  

\end{abstract}


\section{Introduction}		\label{intro}

Beyond the exchanging of voice, image or video information in the fifth generation cellular networks, researchers are exploring new dimensions of interactions including five sense communications, ubiquitous instant communications, virtual/augmented reality (VR/AR), the internet of skills, wireless brain-computer interfaces, holographic communications, and pervasive intelligence, leading a true immersion towards distant environment \cite{zhang2017software, zhang20196g, guo2021enabling}. Concurrently, the evaluation of smart communications in different sectors such as remote learning, autonomous driving, healthcare, industry internet, secured transactions, and smart cities offered by 5G networks inspired experts to research on 6G. 6G wireless networks are envisioned to the provision the future data-intensive smart societies with full automation through seamless integration of wireless networks aspects spreading from ground, air, space to underwater \cite{chowdhury20206g}. With the enormous growth of data traffic and emerging multimedia applications, 6G is envisaged to experience 607 Exabytes by 2025 and 5016 Exabyte by 2030 per month \cite{tariq2020speculative, saad2019vision}. The 6G mobile networks are anticipated to be essentially virtualized, software defined, and cloud-based systems with the motivate to substantial number of ubiquitously connected heterogeneous devices including the internet of everything (IoE), to enable an incredible range of network services \cite{zhang20196g}. With this paradigm shift towards 6G, Internet of Things (IoT) plays a significant role by enabling promising applications connecting the physical devices to the cyberspace of communication world \cite{qi2020integration}.

The advent of future generation wireless networks supports a new set of real time and data intensive applications including sustainable business models, agile management, network orchestration, network slicing, spectrum sharing, and many other vertical services \cite{guo2021enabling}. However, these differentiated services tend to augment a number of issues such as data privacy, network reliability, immutability, security, efficient resource utilization, multiple access control, virtual network functions, etc. \cite{nguyen2020blockchain}. Mobile network operators (MNOs) cannot handle bulky cellular infrastructure effectively especially during the peak traffic demand. This could force cellular operators to raise usage prices for the end mobile users because of the expensive licensing fees for the additional bandwidth and the increasing number of infrastructure deployments and management. Therefore, it is essential for MNOs to alter their existing centralized business model to a flexible decentralized architecture led by software defined networking, network virtualization, and intelligent management \cite{rawat2019fusion, ahokangas2019business}. The decentralized micro-MNOs paradigm demand a scalable and trustworthy solution for enabling quasi real time infrastructure management in addition to improved financial interactions and service level agreements (SLAs).

Blockchain technology (BCT) plays as an immutable distribute ledger for performing secure financial transactions associated with cryptographic symbols. In other words, BCT establishes trusted secure communications among unknown entities in untrusted networks. BCT operated as a common, mutual, and distributed ledger used as a large-scale secure index measuring for all communications architecture particularly in the cyber security domain. BCT has gained momentum for distributed applications and is recognized as the key enabling trusted technologies that received deep attention both in research and industry communities in the context of IoT and 6G mobile communication. Blockchain offers a scalable and distributed platform to record data permanently and validate transactions among different entities without involving central participants \cite{dinh2018untangling}. The integration of blockchain with 6G wireless networks potentially allows MNOs to monitor and manage resource and spectrum utilization with some added advantages like improving spectrum auction, reducing administration expenditures, infrastructure sharing, and so on. Due to the inherited benefit of transparency, BC can record real-time resource utilization and significantly improve spectrum efficiency through the dynamic allocation of spectrum in accordance with the demands \cite{weiss2019application}. With the wireless networks paradigm shifting toward decentralized solutions, plenty of base stations are deployed by MNOs where billions of multimedia devices communicating with each other. As an indication of this trend, fixed spectrum allocation and corresponding resource sharing mechanisms are not effective and scalable as well in the future wireless networks. Implementation of blockchain with the underlying infrastructure not only provides efficient spectrum sharing and resource management but also indulge incentive due to service sharing \cite{maksymyuk2020blockchain}.

Bitcoin and Etherum are the most popular implementations of blockchain augmenting classes of immutable data lists. A decentralized ledger in blockchain technology is used for data recording of each completed transaction among multiple parties on a distributed peer to peer (P2P) networks efficiently \cite{alladi2019blockchain, puthal2018everything}. In addition, BCT provides a trusted, secure, and immutable platform for numerous entities including individuals and industries in order to collaborate, digital asset exchanging, and pertaining transactions with no requirement of controlling by the central authority. Beyond the secure payments and storing financial data, the blockchain applications are expanding to other domains, for example, supply chain management, industrial manufacturing, healthcare, and education sector \cite{taylor2020systematic}. BCT 1.0 generally deals with cryptocurrency (e.g., Bitcoin) and payment, BCT 2.0 handles automated digital finance, and the most recent BCT 3.0 is associated with digital society, for instance, Industry 4.0, industrial IoT, and smart cities \cite{efanov2018all}. The fourth industrial revolution named Industry 4.0 refers to a smart factory setup equipped with intelligent IoT devices. On the other hand, industrial IoT (IIoT) is a subset of IoT that has a narrow focus in transition from conventional industry to smart industry 4.0 ameliorated by the interconnectivity of our digital society and the eminence of IoT devices.

Albeit IoT empowered applications will smooth human life and facilitates smart living, but it is a daunting job for 5G networks to support such advanced applications. With IoT evolving, 6G will overcome the limitations of 5G capabilities supporting ultra-speed data rate, low latency, a wide range of coverage, localization, and other massive machine type communications. The integration of blockchain and IoT technologies handling the limitations with solutions is studied in the literature \cite{panarello2018blockchain} but failed to identify the exact problems to resolving the complexity aspects. Authors \cite{hang2019design} investigated data sensing integrity in an integrated IoT-blockchain platform objective function to enable effortless access to the nodes in a diverse area. However, execution time, communication, and computational overhead issues are ignored. The latency and computational power issues remain unsolved while studying the effectiveness of BC-enabled intelligent IoT systems with artificial intelligence \cite{singh2020blockiotintelligence}. Huang \textit{et.al.} \cite{huang2019survey} surveyed an overview of the green evolution of 6G networks describing architectural transformation and related potential technologies including blockchain. Authors focused on the blockchain based ultra-dense cellular IoT architecture and machine learning approach from the perspectives of research challenges and communication technologies \cite{sharma2019toward}. Reference \cite{hussain2020machine} provided an extensive survey on resource management in cellular networks and IoT systems leveraging machine learning approaches. Although such works have laid a solid background and breakthrough technologies on 5G based IoT, the role of blockchain in 6G enabled massive IoT systems has not been covered in the previous surveys. To fill this gap, this paper surveyed the functions of blockchain in 6G and industrial IoT in terms of network architecture, transparency, advanced network management, key achievement of secured transactions, etc. with a wider range of opportunities. Table \ref{table1} summarize the existing literature surveys and reviews in the context of blockchain enabled wireless networks and IoT systems. All the related papers presented the potential features of blockchain in the context of internet of things pointing to different applications. In contrast to other relevant papers, in this article, we dive deeper in the field of BC enabled industrial IoT and 6G communications with detailed recent trends, challenges, solutions, and future scopes.

\begin{table}
\centering
\caption{Literature reviews on blockchain for IoT and wireless networks} \label{table1}
{\tabulinesep=1mm
\begin{tabu}{|m{13mm}|m{37mm}|m{5mm}|m{94mm}|}
 	\hline
\textbf{Reference} & \textbf{Journals/Book Chapters} & \textbf{Year} & \textbf{Contributions} \\ \hline
 	\cite{fernandez2018review} & IEEE Access & 2018 & Development, deployment, challenges of BC based IoT applications \\\hline
 	\cite{ferrag2018blockchain} & IEEE Internet of Things Journal & 2018 & BC based system performance, computational complexity, communication overhead, security issues for IoT networks \\\hline	
 	\cite{shen2018blockchain} & IEEE Communications Surveys \& Tutorials & 2018 & Review BC based urban sustainability and use cases   \\ \hline
 	\cite{alladi2019blockchain}&  IEEE Access & 2019 & Surveys BC implementation in industrial IoT and industry 4.0 applications \\\hline
 	\cite{salman2018security}&  IEEE Access & 2019 & Blockchain-based security services including auth entication, confidentiality, integrity, privacy, and related challenges \\\hline
 	\cite{fraga2019review}&  IEEE Access & 2019 & Blockchain use cases in automotive industry, opportunities, and threat analysis  \\\hline	
 	\cite{lu2019blockchain}&  IEEE Access & 2019 & BCT in gas and oil industry pointed out efficiency, management framework, and information security \\\hline
 	\cite{yang2019integrated}&  IEEE Communications Surveys \& Tutorials & 2019 & Contemplating blockchain in edge computing enabling resource management, self-organization, computation validity, functions integration \\\hline
 	\cite{xie2019survey}&  IEEE Communications Surveys \& Tutorials & 2019 & BCT in smart cities in terms of decentralization, transparency, democracy, trust-free, automation, and security \\\hline
 	\cite{elisa2018framework}&  Wireless Networks & 2019 & BC for privacy and security preserving of e-government system \\\hline
 	\cite{wei2019convergence}&  IT Professional & 2019 & Reviews the security vulnerabilities in the convergence of IoT and blockchain \\\hline
 	\cite{lage2019blockchain}&  Computer Security Threats & 2019 & Benefits of blockchain in Industry 4.0 automation \\\hline
 	\cite{elmamy2020survey}&  Sustainability & 2020 & Surveys BCT in industry 4.0 by means of privacy, integrity, and multi factor authentications \\\hline
 	\cite{singh2020blockchain}&  Blockchain Technology for Industry 4.0 & 2020 & BC for data management in Industry 4.0 applications \\\hline
 	\cite{sekaran2020survival}&  IEEE Access & 2020 & Demonstrate the integration of BC with IoT for automated operations based on mobile edge computation \\\hline
 	\cite{xu2020blockchain}&  Digital Communications and Networks & 2021 & Discuss the integration of blockchain in 6G in terms of resource management, spectrum sharing, network slicing, device-to-device communication \\\hline
 	\cite{kumari2021amalgamation}&  Computer Communications & 2021 & Review a blockchain based IoT architecture for smart city applications \\\hline
 	\cite{wang2021blockchain}&  National Science Review & 2021 & Present cross network sharing of the blockchain radio access network \\\hline
\end{tabu}}
\end{table}

A limited number of blockchain research projects have been emerged in past few years on industrial IoT and 6G communications, hence, the industry-oriented research is still in infancy. Beyond the features of authentication, accountability, asset tracking, BCT plays a significant role in allowing participant nodes to exchange information mutually in a secured manner in the perspective of smart industries. The COVID’19 pandemic has led the world to adopt for virtual video meetings, entertainment, live video communications ranging from education, healthcare to business. Experts have anticipated that the performance demand for real-time and virtual interactions, the escalation of several data and computation intensive verticals, for instance, digital transactions using cryptocurrencies, industrial automation, augmented reality, virtual reality, autonomous driving, and device-to-device communications (D2D) will increase dramatically after post-pandemic age. A number of significant challenges are identifiable in the area of 6G communications and massive IoT implementation to cater envisaged demand surge. The distributed blockchain ledger technology is the key enablers to address limitations and facilitates operational standards. To the best knowledge of the authors, no literature review is conducted in the context of BCT applications in industrial IoT and 6G networks perspectives. The key contributions of this survey paper can be summarized as follows.

\begin{itemize}

\item Demonstrate the potential key features of blockchain technology of future sixth generation wireless networks and industrial internet of things to shed light in the research communities. 

\item Discuss an ecosystem focused BC enabled 6G business model for the sustainable global society, IIoT, industry 4.0 and beyond applications highlighting financial stability, technology transfer, ecological balance and reconstruction, environment conservation, and democratization of government bodies, and innovation and entrepreneurship.    

\item Outline the fundamental implications of distributed ledger technologies for a wide range of 6G ecosystem centered business model transformations of multiple stakeholders e.g., resource brokers, micro-operators, and edge cloud operators addressing scalability, sustainability, and replicability consequences. 

\item Provide extensive surveys on integrity, security, and privacy issues against cyber threats or unauthorized access. After that, necessary steps depend on various exhaustive layer-wise security risks and extent efficient solutions are contemplated in light of adaptive consensus algorithms and cryptographic mechanisms. 

\item The challenging issues encountered and the potential mitigation techniques as well as future megatrends toward BC empowered 6G cellular and industry applications are extensively pointed out.

\end{itemize}

The rest of the paper is organized as follows. Section \ref{sec2} presents the fundamentals of blockchain technology, architectures, merits and demerits, a broad range of application scenarios, challenges encountered, and mitigation techniques. The benefits of blockchain in 6G including technical model, BC empowered decentralized network design, and socio-economic sustainable business model is extensively discussed in Section \ref{sec3}. The convergence of blockchain of IoT systems is presented in section \ref{sec4}. Section \ref{sec5} demonstrates the security and privacy issues. Furthermore, the summary of this survey and the research challenges, future research directions, and lessons learned are presented in Section \ref{sec6}. Finally, Section \ref{sec7} concludes the review paper.

\section{Background}		\label{sec2}

\subsection{Overview of Blockchain Technology (BCT)}

In recent years, BCT has drawn momentum as an emerging technology in both academia and industry for distributed applications in 6G and industrial IoT applications over the traditional centralized system. Blockchain is a distributed public ledger technology that circulates across all the nodes in a peer-to-peer (P2P) network in order to record and verify transactions between participants without involvement of the central trusted party, where a ledger database maintaining a growing list of information blocks securely employing public key cryptography \cite{jiang2021road}. The fundamental features of BCT include transparency, immutability, security, anonymity, and the least processing time. With the added merits of economic and legislative features, blockchain has the potential to disrupt mobile network business enabling dynamic mobile network operator (MNO) switching by end entities, quality enhancement, borderless roaming, and so on in 6G networks. In blockchain based approach, IoT nodes can be classified into two categories, full node and light node, depending on computing and storage capability. The data collected by intelligent IoT devices combing with the computing resources from nodes enables the distributed IoT providing services to all participants in the entire system.

\subsection{Blockchain Architecture}

A new block is identified by a hash value using secure hash cryptographic algorithm 256 \cite{elisa2018framework} according to the information of its parent block where the hash value of current block (parent) is linked and stored in the subsequent block (child). Note that the hash value changes in accordance with the alteration of block contents and the updated information are spreading throughout all participants in order to invalidate that block. However, the nodes have secret keys assigned for digital signature and validation of the transactions made by the entire mining participants in the network. Thereby, BCT does not need to involve a trusted third party or any central intermediary authority in the BC architecture and hence, it minimizes the processing time and cost while performing P2P transactions. 

\begin{figure}
  \centering
  \includegraphics [width=0.9\columnwidth] {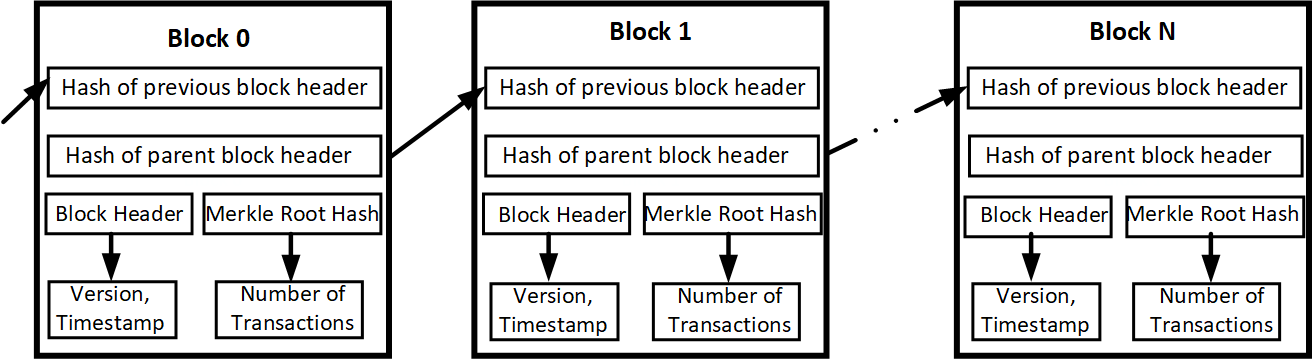}\\
  \caption{Illustration of BC ledger with block details.}\label{fig1}
\end{figure}

\begin{figure}
  \centering
  \includegraphics [width=0.7\columnwidth] {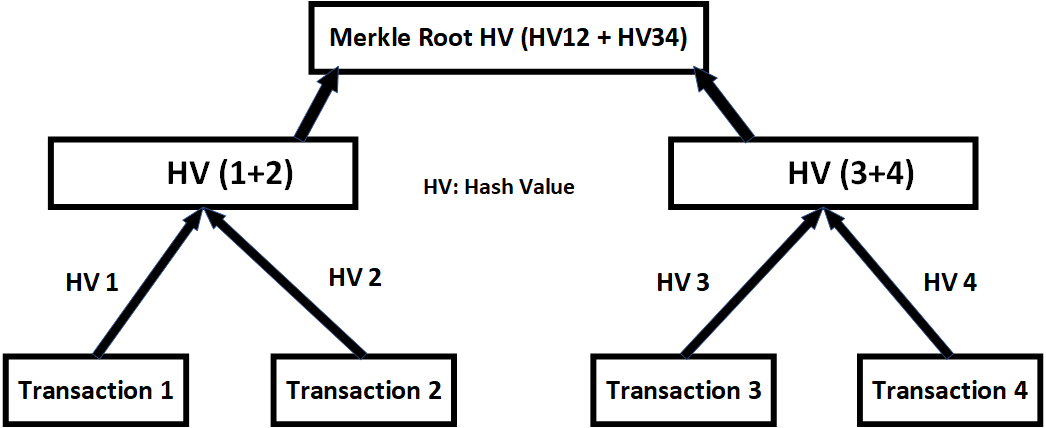}\\
  \caption{Illustration of concatenated Merkle tree.}\label{fig2}
\end{figure}

Each block has only a single parent and child that consists of header information and transaction data. The header primarily comprised a set of following information such as a number of transactions incurred, timestamp, block version, proof of difficulty, and the hash of its parent block, and concatenated hash values of all transactions in the chain. On the other hand, a chain contains many connected blocks continuously grows as BC users to perform transactions. A number of transactions are concatenated into the blocks employing the Merkle root hash tree. The first block called the genesis block from where a chain of blocks is originated as shown in Fig. 1 \ref{fig1}. A new block is connected to the blockchain through a hash value depending on the information of its parent block. A validated block is appended at the end of BC by the reference implying the parent block. A valid block may contain two or more children blocks, when two or more nodes are connected to a block simultaneously generating multiple branches from the same parent. However, timestamp records current time and Merkle root listed hash values of all transactions linked together in the entire block. Fig. \ref{fig2} presents the formation of Merkle tree a pair of transactions are hashed recursively till to form one root at the top of the tree.

\subsection{Features and Potentials of Blockchain Technology}

A distributed ledger has the knowledge of network state information and legitimates any necessary changes to be tracked. Blockchain has proven a trusted effective key enabling technology by means of scalability, interoperability, security, and privacy in data communication. In addition, BC enables the digitalization of real world assets using cryptocurrency and exchanges them via smart contracts. This subsection discusses the fundamental features of BC that make it attractive for 6G wireless networks and IoT systems. 

\paragraph*{\textbf{Decentralization}}

Blockchain uses numerous consensus algorithms to validate transactions and maintain the ledger in a distributed way without the intervention of central authority (e.g., government, bank). The blocks can be replicated and shared throughout the network in order to avoid the failure risk, ensuring flexibility, and strengthening data persistency. 

\paragraph*{\textbf{Immutability}}

Blockchain does not permit the deletion of confirmed transactions since each block is connected with others via the hash value. Modification of the information or breaking the whole chain is very limited. This feature solves the altercation among multiple nodes (e.g., MNOs) related to national law violations, for example, fraudulent transactions by unwanted users. 

\paragraph*{\textbf{Security and Privacy}}

The nature of asymmetric cryptography in BC, a unique cryptographic has is assigned to each block transaction that guarantees the integrity of the information. The inherent features of consensus algorithm, immutability, and anonymous identification ensure the trustworthiness, privacy, and security of the entire chain without any modification of past transactions.

\paragraph*{\textbf{Transparency}}

All the participant nodes have equal rights to access all the information of transactions in BC. In other words, any nodes can interact with others while the confirmed transactions remain unchanged. However, a restriction is not employed for functioning in the entire system. 

\paragraph*{\textbf{Trust-less}}

The participants are not definite to others in the network. The network entities can converse, assist and team build with others without specialized any digital identity to carry out transactions among them.

BCT primarily include four elementary components. 

\paragraph*{\textit{Ledger}}

It is a distributed and shared database containing the information of all performed transactions within the network. Ledger ensures the verification and validation of every transaction among plenty of entities concurrently. The information cannot be deleted or modified once the transactions are performed successfully. 

\paragraph*{\textit{Consensus}}

Consensus protocol validates each action in the network in order to avoid single miner node dominance in terms of the entire blockchain network management and manipulating the transaction history. However, the participants have to reach an agreement using predefined consensus blockchain algorithms. Proof of Stake (PoS), Delegated PoS, Proof of Work (PoW), Proof of Data (PoD), Proof of Authenticity (PoA), Proof of Existence (PoE), and practical Byzantine-Fault-Tolerance (PBFT) \cite{alladi2019blockchain} are the different consensus mechanisms contemplated to validate transactions. For instance, PoW is used in Bitcoin whereas Bitshare and Ethureum employ PoS and DPoS \cite{elisa2018framework}. A new block called mining is created under PoW protocol, which requires to solve a cryptographic puzzle for each individual block. The winning miner who solves the puzzle first and generates new blocks are rewarded using cryptocurrency. The miner nodes add a new block (mine) in order to solve a complicated mathematical puzzle that demands significant computational power and thereafter, gets rewards in bitcoins \cite{elisa2018framework}. 

\begin{table}
\centering
\caption{Comparison of different consensus algorithms.} \label{table2}
{\tabulinesep=1mm
\begin{tabu}{|l|c|c|c|c|c|c|}
 	\hline
\textbf{Algorithm} & \textbf{PoW} & \textbf{PoS} & \textbf{PoW+PoS} & \textbf{PBFT} & \textbf{DPoS} & \textbf{PoA}  \\ \hline
 	Transaction/sec & 10 & 10 & 100 & 1000 & 1000 & 10000 \\\hline
 Latency & 10-60 min & 5-60 min & 5-50 min & <1 sec & <1 sec & <1 sec \\\hline		
  Security & High & Low & Medium & Low & Low & High \\\hline
   Scalability & Low & Low & Low & High & High & High \\\hline
   Throughput & Low & Low & Low & High & High & High \\\hline  
   Power consumption & High & Low & Medium & Low & Low & Low \\\hline
   Fairness & Medium & Medium & Medium & Low & Low & Low \\\hline
   Anonymity & High & Medium & Medium & Low & High & Low \\\hline
    Type & \multicolumn{3}{|c|} {Permissionless}  &\multicolumn{3}{|c|} {Permissioned} \\\hline
   Block Creation & \multicolumn{3}{|c|} {Probabilistic}  & \multicolumn{3}{|c|} {Deterministic}  \\\hline
\end{tabu}}
\end{table}

The node possessing more computational and processing power has more chances to be the winning manner. PoW has a feature that any user can build a block after proving a specified amount of work. However, PoW is an expansive consensus process that requires considerable computational resources. In PoS, a node generates a new block deterministically based on its wealth only to corroborate new transactions and blocks. Unlike PoS, the PoW consensus protocol does not depend on the mining process but execute block validation. Each validator who executes block validation owns a security deposit stake in the network called a bond. In contrast, DPoS employs real time voting and reputation scheme to solve consensus problems by creating a board of limited trusted delegates for validation. Finally, in PBFT approach, all the participants have known each other and can select the priority of importance \cite{cao2020performance}. It is worth mentioning that the blockchain performance highly depends on a suitable consensus algorithm since MNO experiences thousands of transactions per second based on user mobility, traffic intensity, and other factors. However, security, scalability, and throughput are three primary concerns among the aforementioned algorithms. Considering the dynamic deployment scenarios, new consensus protocols are expected to be conceived in the foreseeable future. Table \ref{table2} shows the comparison among various consensus algorithms adopted in BCT \cite{xiao2020survey}.

\paragraph*{\textit{Cryptography}}

This feature secured all network information using asymmetric encryption and allowed the authorized entities to decrypt the original data.

\paragraph*{\textit{Smart Contracts}}

This feature allows software-defined algorithm of distinctive transactions automatically with self-executing computations in accordance with the predefined conditions. In other words, a smart contract is an automated digitalized transaction protocol that executed the contents and terms involved in a multi-step and distributed way. Moreover, smart contracts can empower the potential of software defined networks in 6G networks for enhanced network management and flexible agreements between end users and cellular operators \cite{rawat2019fusion}. There have multiple use cases and applications of smart contracts for facilitating distributed functions such as voting, auctions, crowdfunding, micropayments, etc. \cite{yrjola2020could}. All peer nodes in the P2P blockchain network can inspect the operations of smart contracts cryptographically using computerized transaction protocol. 	

Based on the accessing strategy and permission of entities to write BCT network, BCT can be classified into four types \cite{elmamy2020survey}. 

\paragraph*{\textit{Permission less}}

Everyone with computing power participates in public BCTs to perform transactions and join in the mining node and consensus process for adding new blocks of transactions in the chain. This type of BCT normally used PoS or PoW consensus algorithms \cite{fernandez2018review}.

\paragraph*{\textit{Permissioned}}

Only the approved users can participate for specific business needs in the scheme. Approved organizations are likely to have connectivity with existing applications.

\paragraph*{\textit{Public}}

Limited trusted entities can access the BCT network and be able to write/modify the data, where entities may opt for consortium BCTs. The public access chain is a permissionless proof based security scheme, which assures network reliability for every user regardless of their identity at entry level access. The public chains essentially bring significant benefits for adhoc networks where the restriction of identification is broken for panoptic data exchange. In addition, public chains can substantially regulate the order of participants and increase the efficiency of the community \cite{zhang20196g}. 

\paragraph*{\textit{Private}}

Only the specific users have the permission to write over BCT is given to a central body. Private BCT offers flexibility by increasing control over the transaction rules, which can be changed by overall consensus. The private/consortium is permissioned where the entry of the participants is controlled but not kept in secret. The private chain has a stable community composition and network experiences a smaller number of security threats from unknown attacks \cite{zhang20196g}.  

\begin{figure}
  \centering
  \includegraphics [width=0.7\columnwidth, height = 90mm] {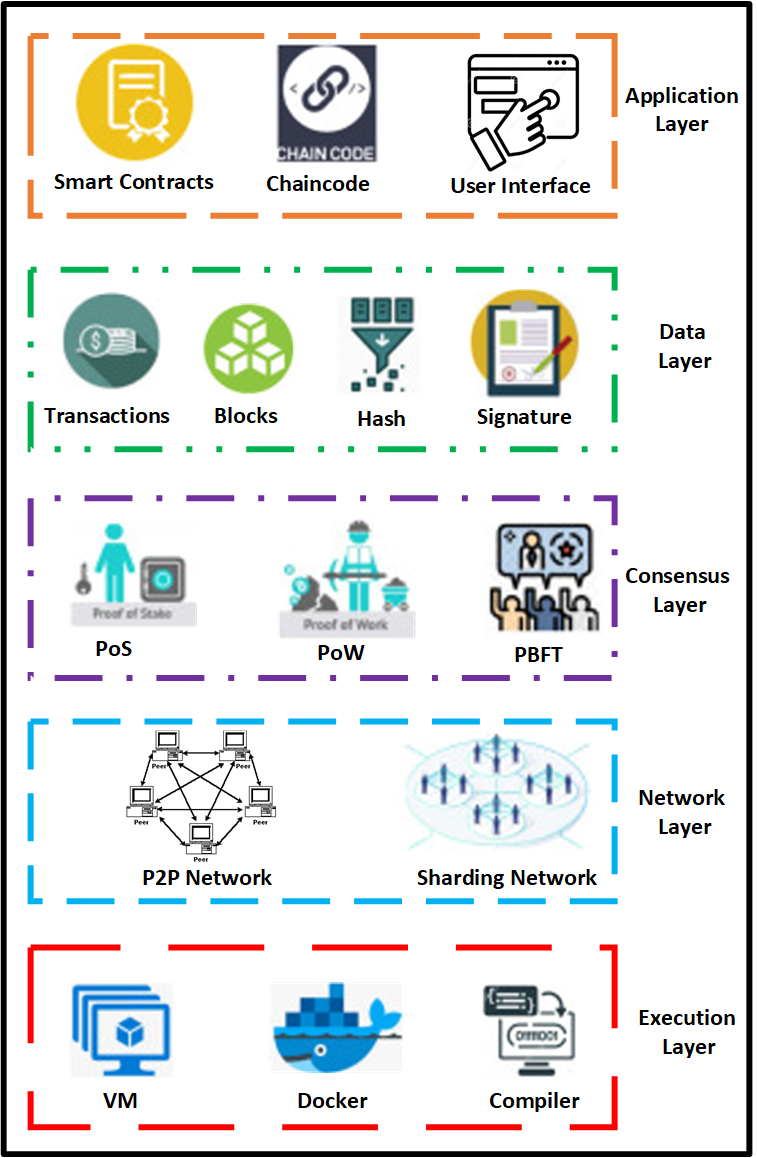}\\
  \caption{Layered structure of Blockchain.}\label{fig3}
\end{figure}

Network layer broadcasts transactions and control messages to the entire network. Each participant fetches unprocessed transactions for newly mined blocks that will require scaling to downsize the number of incumbent transactions. However, the network topology and the broadcast protocol can be reconfigured to be more efficient and dedicated for the high-speed relay network. Consensus layer assures three way trade-off among consensus speed, bandwidth, and security for all participants. Sharding is the potential extent efficient option that divides the consensus tasks among a set of users in order to enhance consensus speed. Storage layer is the leger that stores participants’ state changes resulting from read and write operations. On the other hand, the view layer stores the current state of all nodes and smart contracts, connected to the storage layer. The off-chain with cryptographic proof method curtails the computational burden, and thereby increases system efficiency \cite{nguyen2020privacy}. The operation of transactions, authentication with blockchain, and final execution are performed in the side layer. A details of BCT layered architecture is demonstrated in Fig. \ref{fig3}. 

\subsection{Challenges and Solution Outlines}

Apart from the aforementioned features, blockchain based IoT and 6G wireless systems offer some potential advantages. 

\paragraph*{\textit{Greater reliability and scalability}}

Data transactions are stored in different servers which cannot be altered unless the consent of all active participants through proper consensus protocol. The BC enabled system can readily scale up allowing new users and devices to be connected automatically following the consensus policy. 

\paragraph*{\textit{Improved resiliency and auditability}}

The system is resilient to DoS, DDoS, malware attacks by avoiding single point failure. All the transactions history can be traced easily after operations since data remain unchanged in the system.

\paragraph*{\textit{Improved accessibility and verifiability}}

Information ownership belongs to each individual where they are responsible to validate a new entity to access their information. Storing information at diverse locations increases speedy access easily. However, all new data transactions are verified by all active participant nodes before being added to the blockchain.

\paragraph*{\textit{Enhanced transparency and data quality}}

All recorded data and new transactions are validated prior to making the information authentic with desired quality. All the nodes share the same copy in the network and new transactions are performed in accordance with the associated consensus mechanism. 

\paragraph*{\textit{Higher efficiency and low cost}}

New users can access recorded data based on the accessibility privilege. New records are distributed to all nodes in the chain. Operational cost is significantly reduced since no middle organization is involved to process transactions.

Albeit the potential benefits of blockchain in 6G and massive IoT applications, numerous challenges required to be addressed prior its widespread deployment. Some implementation challenges and possible solutions of blockchain are discussed following.

\paragraph*{\textit{Scalability}}

Because of its distributed nature, cryptocurrency such as Bitcoin requires large bandwidth, ample storage and computing capacity to pledge the ledger integrity. Bitcoin protocol deals with a massive number of information being transferred and broadcasted among the participant nodes. Latency and scalability issues restrict the performance of blockchain enabled applications. For instance, Bitcoin has maximum achievable throughput of seven transactions per second of a 10-minute latency for a confirmed block and four days bootstrap time. In contrast, Visa transaction promises about 56000 transactions/ sec \cite{ling2019blockchain}. 

\paragraph*{\textit{Double spending and other attacks}}

A user perform two distinct cryptocurrency transactions from the same amount of currency, thereby, it breaks the integrity of the distributed ledger. However, a miner may damage honesty of miner blocks in PoW consensus mechanism to achieve more reward under selfish mining attacks. On the other hand, a user generates multiple BC accounts to leak transaction privacy and manipulating transactions. 

Different cryptographic algorithms can be employed to prevent disclosing users’ private information among participants. 

\paragraph*{\textit{Ring Signature}}

A digital signature is applicable for a group participant where all users remain anonymous and ensure that the message is endorsed by the members. A privacy-oriented cryptocurrency called Monero uses ring signature to guarantee the confidentiality of members’ private information \cite{nguyen2020privacy}. 

\paragraph*{\textit{Zero Knowledge Proof}}

This method permits honest entities to convince other parties of statement validation with high degree of confidentiality. The downside is the increasing complexity of transactions approval. Zcash, a privacy protecting cryptocurrency is the example of zero knowledge arguments in BCT \cite{nguyen2020privacy}. 

\paragraph*{\textit{Coin Mixing}}

This method makes obscure the coin owners’ address to assure their anonymity through shuffling the information and coin transferring. CoinShuffle is a decentralized coin mixing scheme, where each node generates an output address before transmitting the information to other participants. When all the nodes have their own output address, they broadcasted to the whole network. However, the computation overload and bandwidth usage are further increased in this policy. 

\subsection{Blockchain in Industry 4.0 and Industrial IoT}

The concept of Industry 4.0 is expected to lead distributed, dynamic, and automated production networks driven by multiple key technological enablers such as blockchain, industrial IoT, artificial intelligence, edge computing, big data, human web integration, robotics automation, and open source software. It is anticipated that the ICT sector plays significant role in sustainable industry automation to support global socio-economic sustainability. The interconnected cyber physical systems (CPS) enabled industrial automation in Industry 4.0; thereby, allowing the transformation of autonomous and dynamic systems from the conventional industrial production process and infrastructure \cite{elmamy2020survey}. In other words, BCT helps to build sustainable development by providing safe and secure communication schemes using key authentication to assure access control management, nonrepudiation, and authenticity. All the smart entities communicate autonomously with each other to attain the common goal under such a highly integrated network. Industry 4.0 is based on the idea of a smart contract in blockchain, which plays a key role of protecting service level agreement (SLA) specification between a supplier and consumers from violation by enforcing trustability and data integrity \cite{hang2019sla}. SLA service is integrated with a smart contract among different participants to satisfy SLA specifications; thus, blockchain technology facilitates SLA services in Industry 4.0 applications with a greater level of integrity and traceability. 

The realization of Industry 4.0 encompasses three paradigms with the different functioning processes including cloud computing, big data analytics, logistics, enterprise resource planning, and product development \cite{xu2018industry}. The smart entities control the resources and regulate the industrial manufacturing process till the end. Secondly, the conventional production process is converted into distributed, flexible, self-organizing, and adaptable production lines done by the smart machine in a cyber physical system. An augmented operator is the third paradigm that facilitates the capability and flexibility of a human operator by means of handling, monitoring, verification of the production process in the manufacturing industry. Leveraging BCT provides a cooperative environment for augmenting workers using a human-machine interface collaborating among entities in the whole manufacturing process. For instance, the real-time information sharing among entities via digital communication network of remote edge computing and warehouse data storage capability in a secured way. 

Industrial IoT (IIoT) is a cyber physical system (CPS), where humans, computers, and machines empowering intelligent industrial functioning using advanced data analytics for radical business outcomes. IIoT allows the integration of internet infrastructure, communication protocols, and wireless sensor networks (WSN) to ease the industrial intelligent automated operations for cognition, analysis, and management, which enhance efficiency and safety \cite{alladi2019blockchain}. IIoT architecture primarily consists of three layers namely physical, communication and application layer. The physical layer comprises sensors, manufacturing equipment, actuators, smart nodes, and intelligent data servers. Numerous access network technologies such as 5G, WSN, machine-to-machine (M2M) communication, device-to-device (D2D) connectivity are used in the communication layer enabling services to physical devices in the physical layers for industrial automation and manufacturing. The physical layer and communication layer form an industrial CPS to support the production and development of smart supply chains, smart factories under the application layer. However, the networking, computation, and controlling infrastructures of IIoT CPS enable the intelligent operations of the manufacturing process.  

\section{Blockchain Empowered 6G Wireless Network Architecture}  \label{sec3}

Blockchain is regarded as an emerging technology to unleash the potentials of 6G, enabling distributed network management (e.g., spectrum, caching, computation) systems that can be used for resource sharing and management, data storage, spectrum regulator, and so on. A distributed ledger records and audits the resource trading transactions in a simplified way. Real-time spectrum sharing market using decentralized BC ledger will bring higher spectrum trading efficiencies and nullify spectrum scarcity managing unused relative frequency bands. Blockchain is considered as a resource-hungry technique in terms of data transmission and complex computation. Few research works have been carried out dedicated to BC-enabled spectrum management \cite{dai2019blockchain, qiu2019blockchain, zhou2020blockchain}, but ignored the efficiency of blockchain itself ranging from the framework architecture to the optimization problem. 

The spectrum regulator from the government body builds strong-level policies and constraints on BC enabled smart contracts which are implausible to break. The regulator body maintains fair spectrum regulation throughout the country, harmonizing various wireless access technologies, handling transceiver power and interference, controlling spectrum coordination and trading, and so on. On the other hand, spectrum owners purchase a license from the spectrum regulator in particular frequency bands on a nationwide basis. Depending on smart contracts, spectrum owners can exploit bandwidth or share with other MNOs for better spectrum utilization. In addition, network infrastructure owners can share their resources such as macrocells, microcells, and data servers with other wireless operators relying on smart contracts. This approach can drive new business models where investment by stakeholders will share with existing MNOs contemplating the blockchain platform. For instance, a construction company can build cellular stations for a prospective rental by MNOs. User equipment (UEs) can dynamically switch among MNOs via service level agreement (SLA) in the blockchain. However, end users can act like infrastructure owners optionally by using smartphones for device-device (D2D) communications. End users can select a particular operator among multiple MNOs based on service quality, price of package, volumes, and coverage areas. The tokenization of infrastructure and spectrum is the key aspect of the aforementioned trading over the blockchain. The spectrum regulatory body defined the allocated money price of the radio resource token (RRT) based on the user request, area of deployment, and other economic values \cite{bugar2020techno}. Besides, the infrastructure resource token (IRT) is recognized as a stable coin in fiat currency where the static price is subject to negotiation in BC smart contract based on location, type, energy cost, and so on. Finally, the national cryptocurrency (i.e, USD, EURO, etc.) is digital equivalent to the fiat currency of the respective country where the cellular networks are deployed. Telecom operators have the capability to buy and sell RRT or IRT in exchange of nationwide cryptocurrency via a smart contract platform \cite{bugar2020techno}. 

A national consortium blockchain ledger process the mobile network information to maintain a considerable number of wireless nodes. Whereas a worldwide public BC ledger contains UE information to ensure seamless connectivity over the globe. National consortium spectrum ledger embodies information about system bandwidth, licensing zone, downlink/uplink duplex mode, owner of MNO and assigned MNOs, licensing fees, frequency allocation, and so on. National consortium infrastructure ledger deals with each eNode B (eNB) such as frequency band, positioning, cognitive radio technologies, assigned MNOs, transmit power, coverage area, signal intensity, and so on. National consortium mobile operator ledger consists of information about the service functional area of each MNO, supported SLAs, token balance (RTT/IRT), and so on. Worldwide UE ledger comprises end user’s information, for example, NC balance, uplink data transmission, and tracking area. Notably, the authorized government bodies can access UE information and should not be exposed to the public. On the other hand, MNO can access only particular data that is permitted to ensure enhance service quality. 

\subsection{Spectrum Management}

With the unprecedented growth of high speed data demand of 5G and beyond, the requirement for more spectrum bands also increases accordingly. A dynamic spectrum management policy has been proposed in Reference \cite{li2017channel}, which allows unlicensed participants to access the licensed spectrum opportunistically without interfering with the licensed users. The implementation of blockchain as a trusted database platform has emerged in various spectrum managing aspects such as spectrum auctions, spectrum sensing, spectrum lease mapping, data mining outcomes, and idle spectrum details are recorded securely in the blockchain \cite{liang2020blockchain}. It has been also identified that the features of blockchain bring opportunities to reduce the administrative expenses related with dynamic spectrum management and able to make significant improvement over the traditional spectrum management approaches. Moreover, blockchain helps to overcome the incentive conjoined with dynamic spectrum management and the security challenges in a decentralized manner. A few other advantages of spectrum management recording information are presented here.

In contrast to the conventional middleman database, BC enables participants to have direct control of the recorded data, hence guaranteeing the accuracy of the information including underutilized spectrum, time, interference protection requirements. A smart contract in blockchain ensures access fairness over the carrier sensing multiple access (CSMA) technique where the access history is not coordinated and recorded. Enhanced spectrum utilization efficiency of both primary and secondary users adjusting traffic demands achieved through BC enabled efficient management scheme. The characteristic of BC empowered dynamic spectrum resource allocation prevents fraud by providing transparency, restrict unauthorized users from accessing the allocated spectrum, and pledge the auction payments since all the executed transactions are validated before recoding in the blockchain database \cite{qiu2019blockchain}.

\subsection{Benefits of BCT in 6G}

\paragraph*{Infrastructure and Resource Management}

\begin{figure}
  \centering
  \includegraphics [width=0.8\columnwidth, height = 90mm] {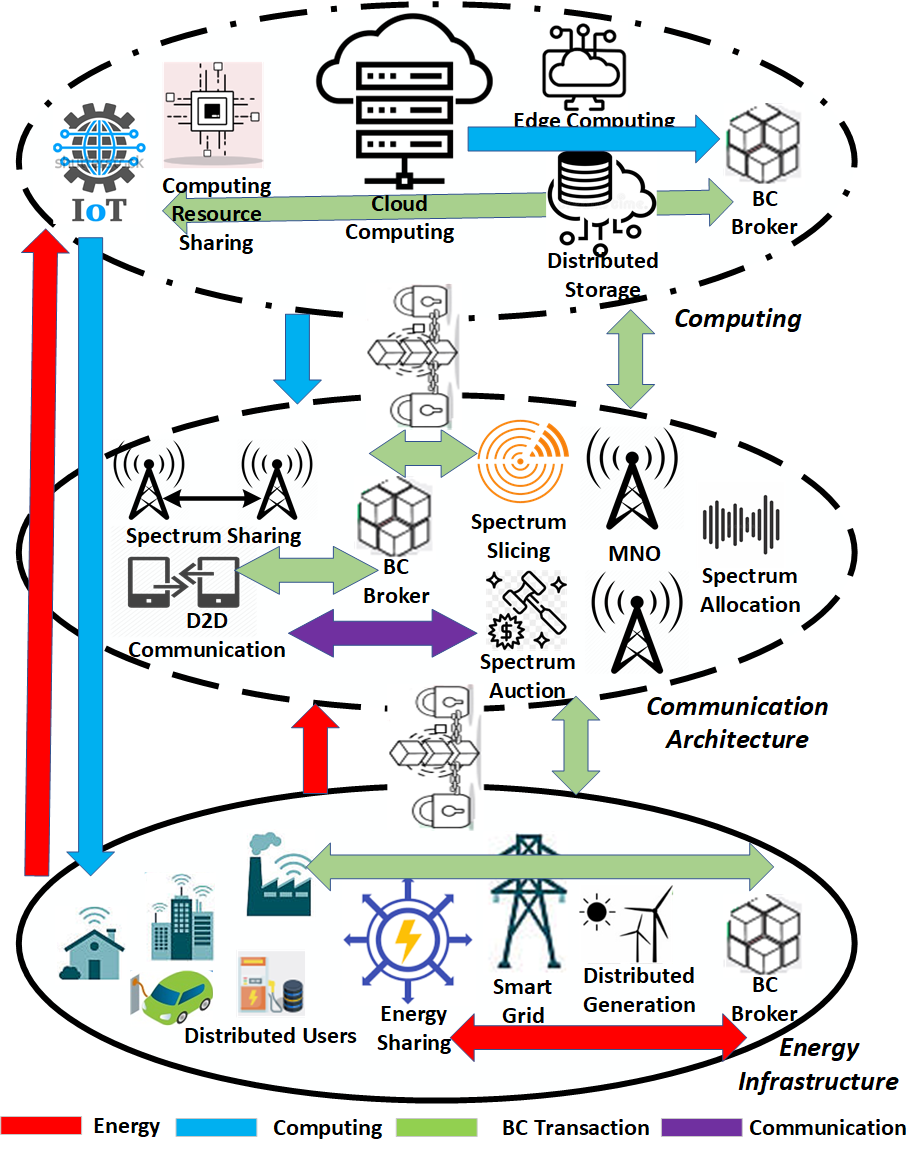}\\
  \caption{Blockchain enabled resource management framework in 6G communications.}\label{fig4}
\end{figure}

The resource management functions including spectrum sharing, orchestration, computing, communication, caching, and decentralization are challenging tasks in the massive connectivity demands of next-generation telecom infrastructure. Blockchain technology solves infrastructure sharing challenges between numerous mobile network operators (MNOs) in multiple ways. Firstly, tracking the infrastructure usage of each MNO and manage the corresponding payment using cryptocurrencies. Afterthat, infrastructure resources can be booked in advance using smart contracts. In addition, additional expenses (i.e., internet provider, electricity usage, real state, etc.) can be equally distributed of infrastructure utilization between MNOs allowing the integration of third parties in the blockchain \cite{nguyen2020blockchain, nour2019blockchain}. On the other hand, telecom regulatory bodies cannot track real-time spectrum usage and restrain unfair behavior under existing technology. Moreover, the mixing of different types (e.g., licensed, unlicensed) of spectrum band from different network operators and wireless standard introduce additional complexity. Furthermore, the complexity of interference handling and real time agreements escalate incentive issues for dynamic spectrum sharing. The integration of automated spectrum trading mechanism under blockchain technology with the existing mobile network infrastructure address the aforementioned problem proving smart contracts agreement among MNOs \cite{dai2019blockchain, qiu2019blockchain}. A details illustration of resource management procedure comprising energy infrastructure, communication and computing architectures using blockchain is demonstrated in Fig. \ref{fig4}.

\paragraph*{Network Slicing and Resource Brokerage}

The smart contract orders the network slice orchestration relying on the service level agreement (SLA) from the broker under the blockchain based network slice trading. A smart contract keeps the record of the usage of each resource and the performance of the service provider against the SLA. The trustworthiness feature enabled by BCT diminishes the collaboration barrier, thereby, ensure efficient ecosystem and security. Moreover, a BC based approach is used to automate the payment, billing, leasing between service providers, and reconciliation in diverse geographies while ignoring the presence of third party in the service layer. Moreover, BCT based automated resource brokerage of network slicing provides some additional benefits, for instance, significant reduction of the operational cost, increase operational efficiency for each network slice, faster the slicing negotiation process and agreement, uplift the enforcement of straightforward agreement related to brokering operations. In addition, blockchain technology also facilitates cross-carrier payment and money transfer across countries, manage license spectrum access, easy provision of roaming terms and agreement among MNOs, and authorization of network elements.   

\paragraph*{Authentication, Access Control, Integrity and Accountability}

Typically, a mobile user uses a subscriber identity module (SIM) in existing mobile networks. Whenever a subscriber needs to change MNOs, one must change SIM card either physically or electronically. A unique wallet ID is assigned to all end users for identification among all MNOs that enable operator-agnostic user quality of experience governed by the competitive mobile network environment \cite{nguyen2020blockchain}. The dispersed ledger technology employed authority, inspection, and complex security for content-centric 6G wireless networks which remain as immutable and transparent for each event that can be utilized during auditing.

\paragraph*{International Roaming and Service Level Agreement (SLA)}

Blockchain mostly depends on the internet connection availability, does not rely on the administrative borders of countries. However, the conventional international transaction method is significantly slow and expensive, while financial transactions using cryptocurrencies across the globe is a matter of few seconds with negligible costs. Therefore, with the advent feature of borderless networking, BC eliminates the international roaming constraint as there is no complexity in data connection between domestic and overseas MNOs \cite{refaey2019blockchain}. It is widely observed that the majority of end users are not entirely satisfied with the package plan because it may be either costly but redundant or cheap but limited data availability. Selecting MNO and an appropriate mobile package plan include a trade-off between price plan, signal coverage, quality of service, and multiple other aspects. Blockchain can address this shortcoming by enabling flexible service level agreement smart contracts between end subscriber and MNO in quasi real time \cite{hewa2020role}.

\paragraph*{Local Area Wireless Networks}

Under the current centralized model, very few cellular operators deploy nationwide wireless networks, experience complex infrastructure development and maintenance. Owing to expensive license spectrum fees, the micro size local MNOs cannot compete in the market in a large scale that leads local telecom operators unprofitable. Blockchain smart contracts address local spectrum allocation problem in real time where a small MNO acquire unused frequency band from large operators over the specified coverage area and time. In addition, blockchain offers greater level policies and monitoring spectrum usage by MNOs across the country, to the central regulatory commission. This feature allows a hybrid spectrum distribution market, where MNOs have the authority to trade parts of spectrum and the regulatory body may restrict bandwidth usage in particular areas, for example, military, border defense to avoid interference.

\paragraph*{Cloud/Edge Computing}

Cloud computing facilitates the offloading of complex computations to distant servers for the resource-constrained user equipment. Edge computing overcomes the heavy computation burden on backhaul networks, long latency, and privacy compared to cloud offloading \cite{dai2019blockchain}. Notably, trustworthiness and security are the primary concerns since sensitive information may involve while offloading computations. Blockchain can build the desired trust between the edge servers and smart devices guaranteeing the integrity of the remote resources and offloaded computations \cite{porambage2019sec}.

\subsection{Ecosystem Focused 6G Business Model}

\begin{figure}
  \centering
  \includegraphics[width=0.9\columnwidth]{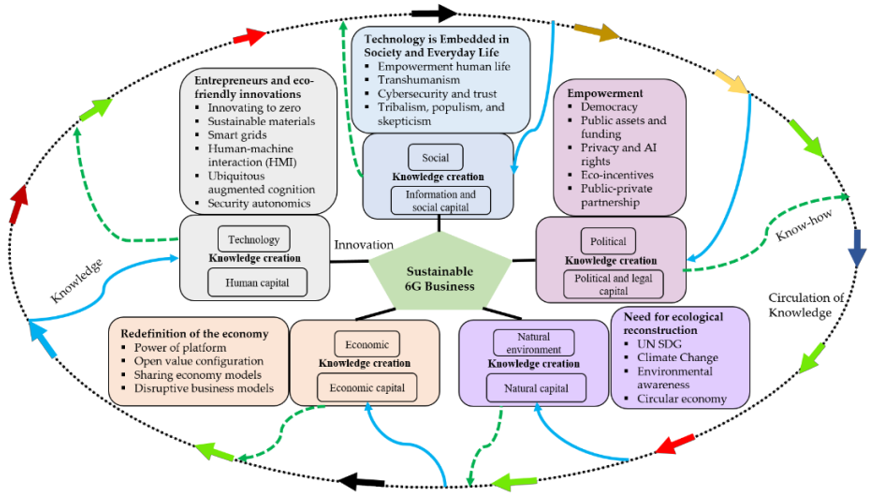}\\
  \caption{Blockchain enabled framework for 6G business model \cite{imoize20216g}.}\label{fig5}
\end{figure}

Given the recent global economic crisis, the business models in 6G will shift toward distributed local edge small operators, software, and virtualized network service providers from monopolized service operators since the existing business structure has a considerable impact on the sustainable global society. Unlike 5G networks, the 6G network offers a wide range of connectivity between real-time dynamic cyber physical world and multiple upfront KPIs, which will inspire various verticals, technology, product, and applications developers for the next generation society 5.0. Moreover, the advent features of 6G tangle with virtualization, intelligence, spectrum sharing between MNOs, edge operations, and a variety of IoT devices; altogether services will push forward into smart living and will be an integral option of smart society \cite{bhat20216g}. Blockchain empowered decentralized architectures envisioned as prospective extension efficient solutions for sustainability, user experience and business, which support multiple verticals in 6G wireless networks perspectives \cite{yrjola2020sustainability}. Sustainability, replicability, and scalability are the proposed key enablers for future 6G business platforms by means of service agility, orchestration, open collaboration, and zero touch management \cite{yrjola2020sustainability}. Sustainability specifies viability, feasibility, environmental, and social impact, whereas replicability refers to external flexibility to comply. In contrast, scalability indicates the potential for internal growth and flexibility. Furthermore, value creation, capture, sharing, and delivery are regarded as key parameters for functioning the 6G business model. Fig. \ref{fig5} depicts the blockchain enabled framework for 6G business model.

It is commonly believed that the future 6G business arena is embodied by an automated and seamless market data collection from dynamic business platforms and humans. 6G will support a user-friendly environment of intelligent big data analysis for sustainable high-end products and services to address specialized demands by the consumers under a wide range of geographic locations. In addition, 6G technology will assist crowdsourcing and advanced cutting-edge distribution to accelerate fair resource distribution in order to support MNOs sharing in sustainable business models. Ecosystem focused sustainable 6G business is key to financial stability, human empowerment driven by knowledge creation, technology transfer and adaption, innovation and entrepreneurship, ecological balance and reconstruction, environment conservation, and democratization of government bodies \cite{imoize20216g}. The advancement of sustainable 6G business solutions will empower IoT devices and other electronic gadgets to operate remotely in a standalone manner and simplify human lives. Peer-to-peer cooperative models and distributed business models under BC enabled 6G solutions is an acceptable approach against a monopoly market that will support a circular economy both in urban and rural areas \cite{bhat20216g}. In addition, BC based 6G business models offer cloud storage for over the top (OTT) companies, cognitive services such as context aware operations, AI, UAV as a service together with calling and data connectivity features, envisioned to attract customers and increase their profit. Moreover, digital twins and the FinTech industry will add a new dimension of business pattern with 6G in addition to the e-government policy regulations. In the case of industry 4.0 and beyond applications, BC empowered business models will provide extreme reliability, privacy-aware secure communications, zero touch assistance, etc. in prolonged lifetime wireless nodes which will bring new business evolution using these technologies \cite{bhat20216g}. As a consequence, the sustainable BC enabled 6G business model will introduce numerous new forms of business structures and opportunities to uplift revenue.

\section{Convergence of Blockchain for Massive IoT }    \label{sec4}

\begin{figure}
  \centering
  \includegraphics[width=0.9\columnwidth]{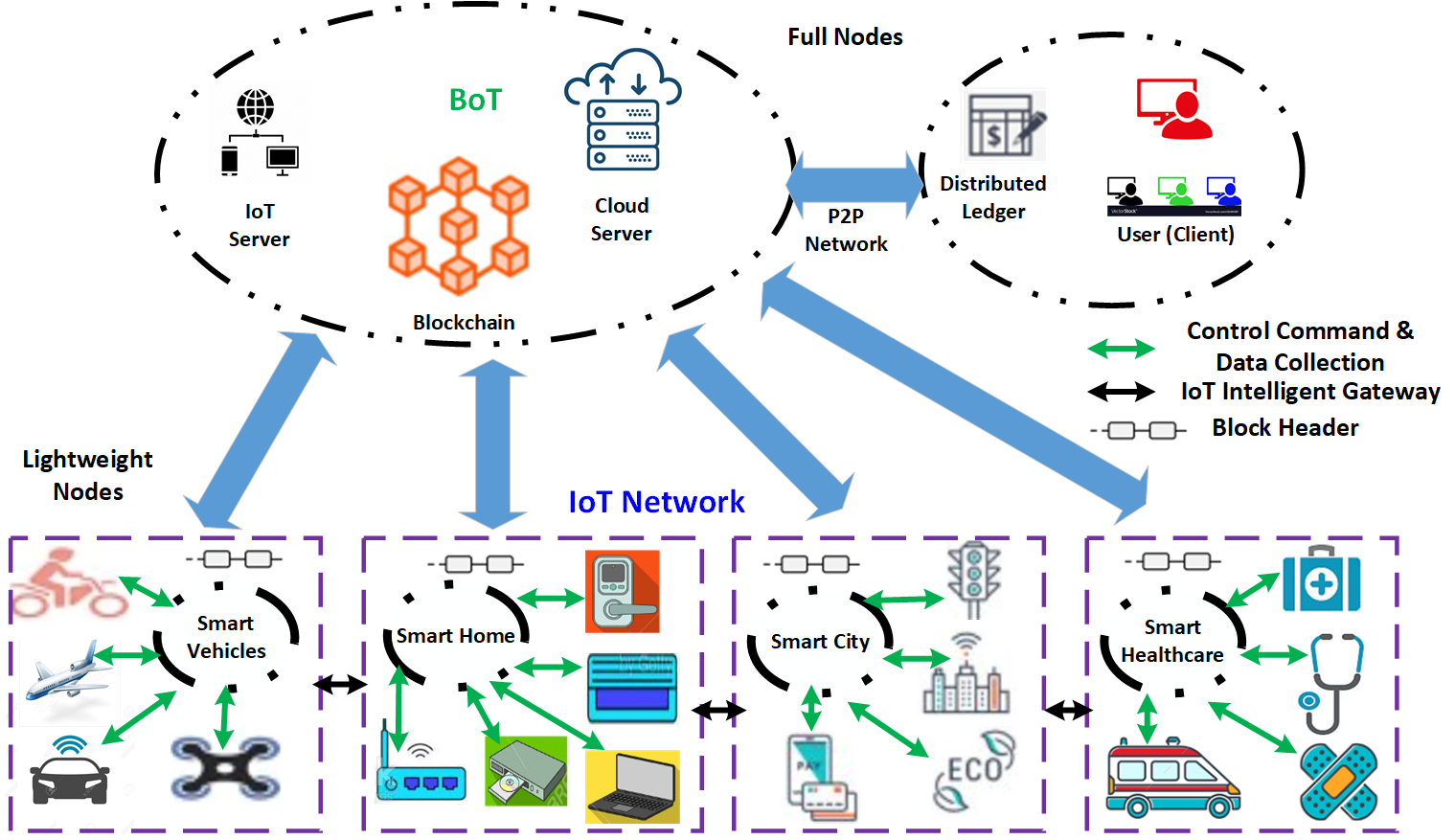}\\
  \caption{BCT based distributed IoT architecture.}\label{fig6}
\end{figure}

In recent years, IoT system integrates multiple distinctive heterogeneous sensors and objects in our smart living era facilitating the information exchange among all nodes. With the rapid expansion of network coverage and intelligent evolution of smart devices, conventional IoT systems failed to meet advanced efficacy, interoperability, network management, privacy, and security vulnerabilities, particularly in the high degree of complex data handling. The centralized IoT architecture experiences high maintenance cost, is vulnerable to unwanted attacks and suffers scalability problems \cite{wei2019convergence, hassan2019current}. To address these challenges, an autonomous distributed system is more feasible for trusted users to participate independently. Moreover, multi participants cooperation ensures the system consistency while a node is crashed by single point failure. A decentralized IoT scheme also reduces the storage load and computational complexity while participants synchronize the entire system copying blockchain leader. The autonomic interaction of IoT devices attained through smart contracts (when the contract clauses are satisfied) in BC that allowed cooperation among each other without any middleman. A decentralized blockchain-IoT (BIoT) architecture is shown in fig. \ref{fig6} illustrating on featured applications.

\begin{table}
\centering
\caption{Potential contributions and challenges of blockchain enabled IoT system in 6G} \label{table3}
{\tabulinesep=1mm
\begin{tabu}{|m{13mm}|m{47mm}|m{45mm}|m{5mm}|}
 	\hline
\textbf{Reference} & \textbf{Contributions} & \textbf{Limitations} & \textbf{Year} \\ \hline
 	\cite{jesus2018survey} & Fundamental functions of blockchain in IoT privacy issues & High computational cost & 2018 \\\hline
 	\cite{qu2018blockchain} & BC structure identify the connectivity between BC and IoT devices for verification & Poor security issue & 2018 \\\hline	
 	\cite{panarello2018blockchain} & Examined the integration of IoT and blockchain & Computational complexity high & 2018   \\ \hline
 	\cite{elisa2018framework}&  Proposed a decentralized e-government P2P scheme using BCT including privacy issues & Suitable consensus algorithms are not focused & 2018 \\\hline
 	\cite{maroufi2019convergence}&  BC offers solutions for IoT devices handling in a smart way addressing limitations & Reliability level not improved & 2019 \\\hline
 	\cite{hang2019design}&  BC integrated IoT platform investigate data sensing integrity in diverse area of applications & High communication overhead & 2019 \\\hline	
 	\cite{dai2019blockchain}&  Intelligent and secure architecture improves the effectiveness using deep reinforcement learning  & Privacy level not improved & 2019 \\\hline
 	\cite{lee2018towards}&  Logchian for guaranteeing block integrity & Security aspect not addressed & 2019 \\\hline
 	\cite{damianou2019architecture}&  BC base edge computing enhance IoT system performance & Resource optimization not addressed & 2019 \\\hline
 	\cite{mistry2020blockchain}&  Centralized architecture lessens the computational overhead & Computation power high & 2020 \\\hline
 	\cite{choi2019multiple}&  BC-IoT prevent hacking and information infringement  & Authenticity is not validated for assuring security level & 2020 \\\hline
 	\cite{biswal2020authenticating}&  Decentralized and asymmetric cryptographic platform avoid the presence of central authority for communication & Data confidentiality is not improved & 2020 \\\hline
 	\cite{yaqoob2020blockchain}&  BC refines the concept of cybetwin for secure production & Privacy issues ignored & 2020 \\\hline
 	\cite{elmamy2020survey}&  Implementation of BCT in Industry 4.0 for cyber threats analysis & Scalability, confidentiality and latency issues are not presented & 2020 \\\hline
 	\cite{sekaran2020survival}&  Investigate security issues of BC enabled IoT and mobile edge computing & Computational cost not addressed & 2020 \\\hline
\end{tabu}}
\end{table}

\subsection{Research Challenges}

IoT systems are heterogeneous in nature comprises of distinctive machines, sensors, robots, vehicles, and so on, which generate different types (e.g., structured, non-structured and semi-structured) of massive data. Moreover, numerous resources such as caching, communication, computing, sensing, control, and localization need to be considered under resource management challenges. In addition, huge IoT data are challenging to manage in terms of transmission, elaboration, and storage under a data management scheme.

\paragraph*{Interoperability}

The information exchanging among IoT devices and the capability of data communications with physical systems is termed as interoperability, which is accounted as a key challenging concern for IoT in 6G due to multiple protocols and cross-layer architectures. Multiple operational platforms such as radio access operators, caching vendors, edge computing vendors, and so on exist together to provide a broad range of services to diverse types of IoT devices. It is impractical to collaborate with multiple service operators, vendors, industrial sectors, different data centers, and IoT systems under existing network architecture. However, a unified authorization and authentication are needed to combine individual systems of different bodies. Furthermore, the provision of diverse services and valid payment methods to various service providers and associated vendors is required to be audited and ensured without the presence of any trusted entity. 

\paragraph*{Privacy and Security}

It is essential to prevent the disclosure of huge data generated by massive IoT devices without users’ consent, which may include confidential and private information. Owing to the complexity, decentralization, and heterogeneity of IoT systems, the protection of users’ privacy and security is a challenging task. Moreover, the recent trend of integrating cloud technologies (e.g., fog/edge/cloud) with IoT systems empowering diverse applications which is vulnerable if clouds are attacked, breaching participants’ privacy \cite{liu2020blockchain}. Note that the current IoT methods for ensuring security, privacy, and information handling capability in 6G heavily rely on third parties.  

\paragraph*{Standard and Latency}

When a greater number of devices are connected in P2P network, the risk of external attack also increases due to the lack of standard regulation device applications. Guaranteeing the minimum latency is another key challenge for high-speed communications for various IoT devices while operating in a large bandwidth range. Table \ref{table3} summarize the potential contributions and challenges of blockchain enabled IoT system in 6G.

\subsection{Possible Solutions}

\begin{figure}
  \centering
  \includegraphics[width=0.9\columnwidth]{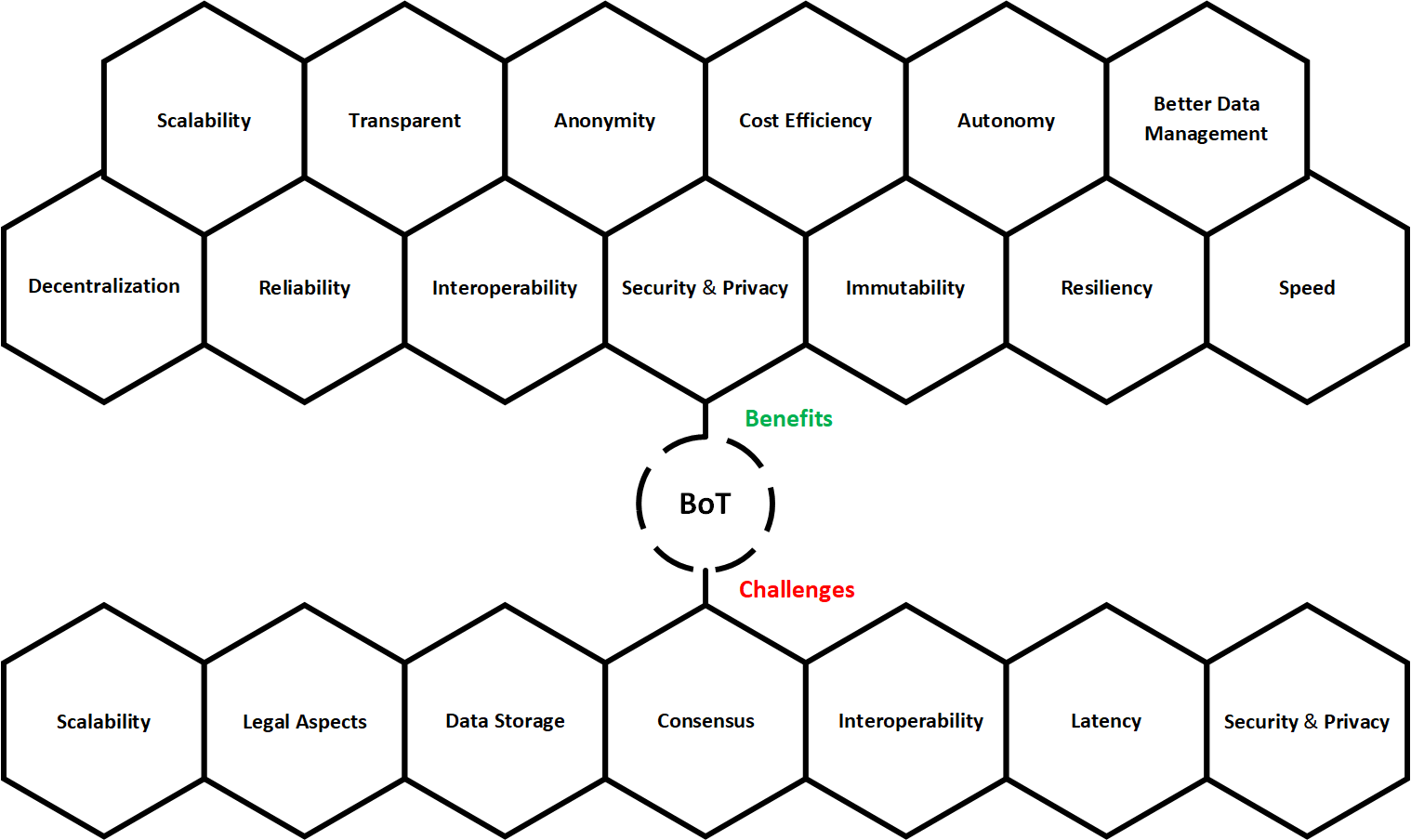}\\
  \caption{Potential challenges and solutions of blockchain-IoT (BoT) scheme.}\label{fig7}
\end{figure}

An integration of IoT in blockchain is termed as Blockchain of Things (BoT). With the inherent emerging properties of BC, i.e., decentralized nature, distributed consensus, cryptographic encryption, trust-less system, and non-repudiation pledge, BoT is contemplated as the key enabler of massive IoT deployment together with the revolution of next generation wireless communication technologies \cite{huang2019survey}. In addition, BoT offers some extended advantages including interoperability, traceability, scalability, reliability, data quality, and trustworthiness. A powerful decentralized BoT architecture is constructed with developer tools for security, smart contracts, and contract execution with a greater level of reliability. BoT can reform the development of IoT with dependence, open and auditable platform. The integration of BoT with 6G offer some potential benefits for IoT deployment and end users, particularly in industrial applications, for example, choosing multiple services from any organization, choosing anything from anywhere, choosing any route from any network. These features allow BoT to provide some extended aspects, such as trust building between 6G enabled IoT applications and end users, cost reduction due to removal of intermediate node and direct service, and real-time rapid transaction service \cite{sekaran2020survival}. Fig. \ref{fig7} presents the possible challenges of blockchain empowered IoT automation with potential solutions.

\paragraph*{Decentralization}

Paradigm shifting towards decentralization eliminating failure to escalation fault tolerance performance. BoT restricts the autarchy of resources against corruption while managing data information \cite{mendki2020blockchain}. 

\paragraph*{Autonomy and Interoperability}

BoT enables autonomy function for implementing industrial IoT applications by means of cooperating mobile devices with each other. Blockchain composite layer on P2P network empower the data communications with physical system and have the capability to exchange information in the entire IoT system with uniform access by providing a unified authorization, authentication, and billing system \cite{ishmaev2020ethical}.

\paragraph*{Traceability and Reliability}

The feature of historic timestamp in blockchain empower the detection and authentication of IoT data (e.g, temporal-spatial information of data block). The trustworthiness of data quality is termed as the reliability of IoT data which can be guaranteed through asymmetric encryption mechanisms, hash inclusions, and digital signature. Each user in the BoT architecture is allowed for validating the transaction legitimacy with full confidence and responsibility.

\paragraph*{Security}

Information exchanges between inter-device communications are regarded as transactions considering new agreements to support security. A manufacturer allows IoT devices for updating securely incorporating secure code deployment.

\section{Security and Privacy Aware BC}     \label{sec5}

Since 6G connects diverse types of voluminous and distributed smart devices, services, and applications, so privacy, security, and reliability are the paramount requirements of data management for high-end users and industries as well. Developing a reliable ubiquitous trustworthy BC model into 6G networks coordinate between end users and network entities to collect and prevent suspicious misbehavior actions including non-repudiation and indirect reciprocity. Despite the potential benefits of BC technology, numerous privacy and security issues against cyber threats or unauthorized access have been key concerns while convergence with 6G networks and IoT devices. It is essential to guarantee the integrity, confidentiality, and availability of data services for each transaction. However, integrity is attained by protecting transaction data without any modification; preventing information to unauthorized users to achieved confidentiality; whilst service availability refers data information is available when needed which is free from denial of service (DoS) attack or other identical service disruption. A DoS attack targets the services or system availability through flooding with requests to its legitimate users to make the service unavailable. In other words, the attacker flooded the server with massive requests pushing the workstation to commit all computer resources and forcing the network server to malfunction leading to a crash. On the other hand, distributed DoS (DDoS) contribute simultaneous flood requests come from multiple sources in the network \cite{elmamy2020survey}.

The buffer overflow within the server disables security systems, especially the firewall and intrusion detection system. In addition, the cloud-based modern intelligent factories and manufacturing production are highly suffered from multiple DoS attacks if the cloud server is not designed to protect against such threat \cite{franke2020survey}. Moreover, the impacts of DoS attack can immensely affect Industry 4.0 applications because of using a substantial range of vulnerable sensors. The fundamental challenge of DoS is the detection and quantifying of the emerging risks in the network. Reports \cite{elmamy2020survey, yan2017industrial} suggest that the transition to industrial IoT (IIoT) develops control policies, builds security awareness and authentication mechanisms including behavioral assessment to prevent hacking and digital encryption technologies. The data flow in IIoT systems manifests innumerable points of vulnerability, and hackers can steal data at any point without suitable protective measures. Particularly, hackers can access cyber physical protection systems which collaborating with surrounding physical systems either physically or logically. With the escalation of collaborative threats and industrial espionage, many smart industries developing intelligent security techniques and control policies to guarantee trust and transparency aiming to defend confidential information \cite{xu2018survey, lu2019blockchain, puthal2018everything}. Using the hash function and cryptography proofs among different linked smart contracts can be regarded as a promising solution in blockchain technology by preserving the integrity and detecting any malicious modifications. Moreover, contemplating a Mongo database in BC can also enforce security algorithm, and offers additional cryptography proofs against confidentiality threats and attacks \cite{elmamy2020survey}.  

Asymmetric encryption and digital signature are employed in the network to pledge security, whereas a private key is assigned to a particular user for validating the transaction, which enables shield for unauthorized access or adversarial attempts to data modification. To increase the privacy of transaction data stored in the system, public key cryptography is used employing incomprehensible hashes of all blocks in the chain. Since the individual user’s data is stored in various nodes in a decentralized P2P system which ensures the service availability ignoring the single point failure. Notably, any valid transactions received from the BC network are permitted by the witness particular for the malicious nodes to prevent unauthorized connections \cite{elisa2018framework}. The different sorts of security services and corresponding necessary steps taken to guarantee adequate security and privacy of transactions are summarized in Table \ref{table4}.

\begin{table}
\centering
\caption{Security issues and common steps \cite{elisa2018framework}}\label{table4}
{\tabulinesep=1mm
\begin{tabu}{|c|c|}
 	\hline
 	\textbf{Services} &  \textbf{Necessary Steps} \\\hline
 	Access control & Digital signature and encryption \\\hline
 	Authentication & BC address and digital signature \\\hline
 	Privacy & Encryption \\\hline
 	Integrity & Encryption \& digital signature \\\hline 
 	Availability & Distributed \\\hline
 	Trust & Encryption \& digital signature, distributed \\ \hline	
\end{tabu}}
\end{table}

The security concerns can be the network and communication security, authentication and identity management, establishment of trust protocol, distributed cooperation, reliable consensus policy, and transaction privacy and security in designing distributed 6G networks and industrial IoT. All the nodes in a distributed IoT system interact autonomously without involving a central server. As a consequence, such openness may lead to numerous impacts, for example, launching text spoofing, node capture, eavesdropping can filch system data partially to compromise the stability issue of massive IoT systems. Compared to RSA based cryptographic algorithm, the Elliptic Curve Digital Signature Algorithm (ECDSA) outperforms in network security particularly for smaller key strings and faster operation \cite{aitzhan2016security}. In order to improve computational efficiency, UE devices will run a lightweight node to store transactions information instead of entire copy of blockchain which leads to a reduction of storage burden. However, the elliptic curve cryptography (ECC) method offers a similar privacy level with Rivest-Shamir-Adleman (RSA) approach, but ECC requires less number of bits for encryption and digital framework. For example, 256-bit ECC provides the same security level as 3072-bit key in RSA \cite{elisa2018framework}. However, a shorter asymmetric key occupies low CPU space, faster key production, and lower memory usage. ECC algorithm has been recognized as a promising solution in terms of safety and reliability in the blockchain context such as Ethereum and Bitcoin. It is worth mentioning that IoT sensing devices are energy constrained and subject to limited storage capacity and computational algorithms. Applying ECC lightweight cryptographic algorithm may fail to meet the demand on limited memory and computational load under energy constrained aspects. According to \cite{wei2019convergence}, ECDSA based algorithm offers optimum communication performance that realizes to balance between memory limit and energy consumption. 

All the transactions containing content, frequency, destinations are shared in the blockchain which may lead to potential security and privacy risks that unwanted entities can infer the real identity of participants. In the convergence of BC and IoE, adversaries can even track physical IP address of wireless nodes and IoT devices which may eventually make system crashes and destabilization through denial-of-service attack. Henceforth, it is essential to protect sensitive information such as physical location, network address, device class, and other private information about the nodes especially during recording the transactions data. A homographic and zero knowledge proof are potential encryption methods for privacy safety via ensuring the confidentiality of transaction records in the public blockchain \cite{lin2018id}. 

The ever-increasing number of IoT devices would demand substantial storage and computing capacity to realize identity authentication. Owing to the high degree of heterogeneity of smart wireless devices, the typical methods in BC may not be able to cope with the requirements of identity authentication in the internet of everything (IoE). To tackle such challenges, a public key infrastructure combing the smart contract with encryption algorithms autonomously handles identity management with a greater level of reliability. However, the reliability can be enhanced by designing device registration including public key, manufacturer, expiry date, identity validation, firmware update, and device obsolescence \cite{hammi2018bubbles}. In addition, synchronization of recorded data and querying the full nodes in the blockchain can effectively implement identity management among all P2P nodes and minimizes the cost of building trust. 

An effective collaboration among participants is essential to accomplish specific functions and services in P2P network, since it is challenging to maintain reliable trust and cooperation in the BC enable IoT systems where each node can enter and leave the system at any time, thereby the identity remains anonymous. The reputation assessment technique is the best way to enhance trust and cooperation in IoT applications such as crowdsourcing and adhoc networks. In contrast, both entity and data-centric trust models suffers privacy leakage, lack of data information, latency, and free riding problems \cite{lu2018privacy}. Establishing node assessment model by extracting the behavioral data of other participants recorded in the blockchain and evaluating the transparent block data. However, the calculation of credibility of each node is done automatically considering their behavior data and timeliness to measure the system performance. Therefore, it is important to account for the incentives and node preferences while choosing the proper behavior type and appropriate time scale to better regulate the node behavior and develop trust. 

\begin{table}
\centering
\caption{Layer based security risks and possible solutions.} \label{table5}
{\tabulinesep=1mm
\begin{tabu}{|m{14mm}|m{29mm}|m{60mm}|m{35mm}|}
 	\hline
\textbf{Layers} & \textbf{Attacks} & \textbf{Challenges} & \textbf{Solutions} \\ \hline
 	Physical & Eavesdropping  & Content modification of a tag & State detection and data recovery schemes   \\
 	& Jamming & Cause exhaustion and radio interference 		&  				\\
 	& Physical damage &   Device connection and recovery &      \\
 	& Firmware replacement  & Condition monitoring &		\\\hline
 Network & De-synchronization & Interruption of existing connection & Hierarchical consensus algorithm  \\
 & denial of service & Reduce the activity of devices & Dynamic committee \\
 & Byzantine attack & Consensus protocol &  \\	\hline
  Transport & Middleman & Breaking integrity and confidentiality & Zero-knowledge proof  \\
  & Message spoofing & Privacy and security & Homomorphic encryption method \\
  & Flooding & Repeating a new connection request and uses frequent HELLO packets until reaches peak level  & Lightweight authentication \\
  & Sybil & Duplicate a single node in multiple locations & \\\hline
   Application & Identity duplication & Authorization and authentication & Customized smart contracts  \\
   & False injection & Sent untrusted data to server & Protocol based cryptographic algorithm \\
   & Malware & Malicious link can attack web and mail services & Reputation assessment model \\
   & Selfish attack & Coalition and trust building &  \\
   & Brute force & Attempt to guess password &  \\ \hline
\end{tabu}}
\end{table}

Owing to the series of problem in the centralized IoT and 6G wireless systems such as inferior scalability, single/multi-point attack, and privacy leakage, blockchain technology offers an immense opportunity for developing secured decentralized infrastructures. Advanced encryption algorithms, reputation assessment models, authentication, and consensus protocols are vital technologies to build the convergence of blockchain, IoT, and 6G networks addressing a high degree of heterogeneity. Table \ref{table5} summarize the layer based security risks and possible solutions in the blockchain domain.

\section{Research Directions and Opportunities}  \label{sec6}

With the rapid development of IoT technology, a significant number of obstacles are incurred to downsize the adoption of IoT devices across many applications. The centralized structure of IoT and 6G networks need the owners to trust the entities to keep data safe. With the potential benefits of blockchain technology, a distributed ledger neglects the challenges of centralized architecture while constructing trust, privacy, and security with guaranteed reliability. Moreover, BCT minimizes the cost by eliminating third party middlemen and intermediary overhead. Blockchain offers many solutions to address 6G and IoT challenges but convergence between technologies create new issues. There are some security and performance issues that exist without solution. For example, the privacy issue of BCT depends on the method and usage of the hardware and software implementation. On the other hand, the increasing number of miners leads to high storage cost, and latency, and lessens the speed of distribution over the entire network. 

\paragraph*{\textbf{Privacy and Security}}

IoT systems and wireless networks are usually suffered from security attacks like eavesdropping attacks and replay attacks \cite{alladi2019blockchain}. In addition to these security attacks, blockchain technology brings their own set of security risks such as message hijack, smart contract vulnerabilities, etc. \cite{li2020survey}. The proactive security policies for dynamic environments will uplift the control signaling cost. However, reactive security techniques are not preferable for 6G networks. Hence, cost effective proactive mechanisms need to be investigated. Privacy leakage is another key concern in transactions and recorded stored data in the blockchain. A massive wireless device connectivity, multi-hop data routing, and heterogeneity in 6G networks manifest the question of data privacy. Numerous adaptive consensus policies need to be designed to handle a massive number of nodes with guaranteed throughput and security \cite{tang2019future}. Outsourcing a wide range of services at self-organization and edge to attain automation introduces new security issues, which urge to be studied furthermore. In addition, vibrant research in the quantum security context seems to be promising. 

\paragraph*{\textbf{Energy Efficiency and Cost effectiveness}}

The current consensus mechanisms are greatly resourced intensive and less efficient, for instance, PoWs consume a great amount of electricity; thus, make computationally expensive \cite{mendling2018blockchains}. Note that more power miners need to be deployed to run heavy algorithms when the dimension of blockchain size is larger.  Proof of trust and DPoS are examples of energy-efficient consensus algorithms store only the recent transaction data instead of recording the entire blockchain data. Nevertheless, resource and energy constrained IoT and wireless nodes are still overwhelmed due to the massive amount of data demand. However, the design of more reliable and efficient consensus protocols is an open issue in order to prevent excessive power consumption and cost overhead.

\paragraph*{\textbf{Resource Limitations and Spectrum Allocation}}

The current blockchain deployments demand stable network connections. Smart IoT devices and wireless nodes are not always ensured reliable network connectivity, therefore, blockchain implementation is a challenging task under this scenario. In addition, large network overhead is another challenge for incorporation into future 6G networks and IIoT applications \cite{dorri2019mof}. Owing to the limited resource allocations, a degree of decentralization in current blockchain implementations 6G and IoT networks are also limited. Due to resource hungry nature of the blockchain system, optimal resource allocation needs to be ensured between the wireless networks and blockchain. However, the profit of cellular operators mostly depends on the volume of offered services instead of the spectrum band leasing option. A limitation of the implementation of BCT in the next generation 6G mobile systems is still a big challenge in the near future.

\paragraph*{\textbf{Regulation}}

Since blockchain operates in a distributed way without the intervention of third-party central bodies, government regulations, industrial and MNO standards need to be enforced. With the augmentation of BC platforms, the necessity of timely enforcement of global standards have increased more \cite{alladi2019blockchain}.  

\paragraph*{\textbf{Machine Learning (ML) enabled BC}}

Real time data prediction and analysis of industrial IoT nodes are important for the successful deployment of BC enabled IIoT applications. Machine learning algorithms can be integrated for data analysis in the node itself to avoid the data transit for prediction and analysis. Notably, ML empowered BCT in industrial IoT applications and 6G cellular architectures is an emerging technology getting deep attention among industry and research communities owing to tackling privacy violations and security issues by preventing data movement. 

\paragraph*{\textbf{Industry 4.0 applications}}

Digital transformation is a promising solution to elevate industrial productivity. Addition of more new technologies could lead to more susceptibility and increasing cyber-attacks intending to manufacture based industry 4.0. A very few works conducted recently creating a dataset to handle cyber security threats in the context of IoT under fog/cloud systems using machine learning and deep learning to develop an adaptive learning framework to address the aforementioned problem \cite{moustafa2019new, 2dai2019blockchain2, mishra2018detailed}. However, integration of BCT based new cutting-edge technology for Industry 4.0 and beyond is a paramount challenge for successful deployment, which is contemplated as a double-edged sword.

\paragraph*{\textbf{AI enabled blockchain}}

Typically, AI methods depend on a centralized architecture which may lead to data tampering and huge communication overhead. Consequently, delaying and erroneous phenomenon incurred while making a decision by AI \cite{salah2019blockchain}. Therefore, decentralized AI is a compelling need to provide reduction of network traffic congestion and privacy preservation avoiding data uploading collected at edge equipment to the central data server. Nonetheless, distributed inference and training still need communications between a central server and distinctive multiple agents for data sharing. As a result, the model updates may be disclosed, and it is difficult to assess the contributions of individual agent and so lacks motivations for data sharing. Distributed AI-based blockchain is considered as an emerging solution replacing the centralized server where data sharing is performed autonomously through smart contracts. To unleash the potentials of two individual disruptive technologies such as artificial intelligence (AI) and blockchain, it is a promising opportunity to combine these megatrends to make mutual enhancement and meet the distributed requirements intelligently. AI brought some potential benefits to blockchain such as AI can manage blockchain operations efficiently and AI can fully utilize the encrypted data. The blockchain operation requires a great amount of computer computation processing power, which makes the system less efficient. For instance, a brute force approach (i.e., hashing algorithm) for mining nodes in Bitcoin attempts characters combination until the verification of the transaction. Moving away from the brute approach, AI enabled smart contracts provides better performance and desire level of quality \cite{qiu2020ai}. The information in a blockchain database held in encrypted state offers security, but experiences challenges for data usage \cite{jaschke2018unsupervised}. AI algorithms address this challenges with further data security and make full use of the BC database. Research in physical layer security, adaptive channel coding, NFV, network slicing, SDN employing BCT is still in the initial stage that requires further investigations. The role of AI with BC to the 6G communications and IoT architectures will make society more efficient and super smart through solving more complex problems that are often changing dynamically. 

\paragraph*{\textbf{Performance enhancement}}

Blockchain is predominantly resource-hungry in forms of caching, computation, transmission, and wireless resources particularly at the edge network for new block generation, validation, ledger storage, consensus policies, etc. BC normally performs as an overlaid layer to serve the underlaid layer in a large scale services. Hence, the performance enhancement of BC enabled systems should be further investigated. 

\paragraph*{\textbf{Optimization}}

IoT based applications are largely in heterogenous forms in terms of various technical requirements, for example, some services require extremely low latency, while some other services concern only about privacy. Thus, blockchain protocols needs to be modified with advanced powerful adaptivity to provide services in 6G networks and IoT systems as well. Notably the properties of BC such as delay and scalability issues limit its own performance. Therefore, a novel consensus mechanism aims to be designed to make the full use of blockchain optimization for enhancing the performance in terms of throughput, security, time to finality and so on to meet the up-trending demands. 

\paragraph*{\textbf{Business Model, Data storage and analytics}}

Internet of everything (IoE) is more general form than IoT, which connect users, data, processes, and things seamlessly in an intelligent manner. With the augmentation of IoE, massive number of things generate real-time data streams continuously. Consequently, sufficient and efficient data storages are required, where blockchain plays an important role for decentralized storage technologies in different domains. However, for the accurate and efficient decision processing, effective data analyzing research methods are required within the distributed BC enabled data storage. With the help of digital technologies, processes are optimized and automated that will re-invent new business models in variety of industries. With the proliferation of business agility and velocity in IoE platforms, it is essential to investigate numerous possibilities of developing modern business models. Furthermore, the impact of blockchain based technologies intending of interoperability functions among multiple businesses needs further research. It is readily agreed that 6G will greatly support sustainable goals (SDGs), which requires mapping and depth analysis of evaluating KPIs and social impact. Extensive research works need to be carried out to bring equality in gender, society, and privacy to support new use cases.

\paragraph*{\textbf{Virtualization of Access Network}}

Although 6G networks are distributed, still an intelligent support are needed in terms of visualization and SDN. The virtualization in 6G requires to compatible with all radio access interfaces including quantum radio, smart surface, THz communications, which is a big obligation to address. The key concern, will every network node support different vertical requirements such as wireless resources, cloud computation, latency, caching and QoS simultaneously?

\paragraph*{\textbf{Edge Computing and Mobility}}

In general, satellite networks generate a significant delay during vertical transmission under aerial, terrestrial and space domains. However, high altitude platforms (HAP) in aerial access networks offer edge caching and data computing facilities minimizing the computations and delay of satellite networks providing critical services. Therefore, edge caching demand further robust research in terms of resource allocation, localization, efficient computing, power efficiency, etc.

A cell-free communication needs precise synchronization and localization in heterogeneous networks (HetNets) with beamforming, which requires further investigation. It is infeasible to ensure high precision beamforming when dealing with high mobility. Hence, location aware mobility tracking for beamforming looks promising. An efficient integrated protocol design is necessary to maintain communication and localization during the mobility among different access networks.

\paragraph*{\textbf{Sustainable 6G Ecosystem}}

Shareable resources are the key assets that are not restricted to inter-domain communication but spread in computing and energy sectors as well, while the communication section also depends on computing resource and energy provision as illustrated in Fig. \ref{fig4}. Nevertheless, a trusted and secured blockchain based trading ecosystem considering communication, computing, and energy can built an efficient sustainable 6G. A distributed energy management is a microgrid system especially in remote areas, where distributed generation (DG) units, for example, PV solar and wind firms are the best example of a sustainable ecosystem approach through distributing electricity to remote localities.  A locally generated distributed energy broke with local providers is a cost-effective approach of trading energy and power line communications instead of connected to the central grid \cite{talapur2018reliable}. The ecosystem is nurtured while the computing and communication relay services for DG in exchange of empowering up the hardware. In terms of the security and performance measurements, a public chain is more preferable to inter-domain blockchain transactions between operators like first tier MNOs and the national grid. However, the private chain can be contemplated for locally oriented resources particularly for IoT and off-grid nodes in remote regions \cite{xu2020blockchain}. To achieve the best results, an ecosystem may add numerous consensuses on different access chains. However, the sustainable business ecosystem is not bounded to the scope of communication, computing, energy, rather it can spread itself to augmented range via cross-field integration, such as financial, automotive, logistics chain, manufacturing, and so on. Beyond the implementation of blockchain, hardware deployment plays a crucial role in the ecosystem to realize portable mobile device solutions like cars, portable IoT devices, and drones.

\paragraph*{\textbf{Future Applications}}

\textit{\\Autonomous Driving and V2V communications:} 

Intelligent vehicle to vehicle (V2V) transportation systems is one of the emerging applications in near future supported technical capabilities under 6G network services. A BC empowered approach develops trust management among smart vehicles, but the limitation of contemplating to adhoc network is the shortcoming. Autonomous driving in 6G offloading from connected autonomous vehicles (CAVs) to edge servers via space networks is demonstrated in Reference \cite{na2020uav}. The implementation of blockchain can store sensitive information for CAVs and audit the offloading steps, which guarantees the security of the system. Autonomous deployment including mobility challenges such as multi junction road intersection requires further investigation. 

\textit{\\UAV and Drone:} 

In recent years, unmanned area vehicles (UAVs) and drones are popularly deployed for civilian usage, diverging from conventional military applications. The incorporation of blockchain in drone industry helps to identity management, air traffic management efficiently with a greater level of accuracy and security \cite{li2018uav}. Integration of blockchain to UAV networks are the potential areas for future exploration particularly in detection of compromised UAVs under surveillance application. If a compromised UAVs sending false information (detecting an intruder) will not be able to join in the blockchain with the help of smart contracts and suitable consensus algorithms. 

\textit{\\Remote Learning:} 

A blockchain enabled platform could record educational databases in a transparent and independent way, thereby, educational authority easily accesses secure records and transcripts. Besides, it also facilitates secure collaboration among universities and other institutions. The inherent characteristics of blockchain allow academic bodies to check educational dishonesty \cite{alladi2019blockchain}. The addition of BC in educational sectors also prevents fraudulent activities and tampering via locating transaction records and verifying the key. 

\textit{\\Power Sectors:} 

With the surge of smart IoT devices ranging from smartphones to smart meters to smart electric vehicles exploiting variable power demands, the electricity grid is becoming more complex to handle with mushrooming distinctive types of power generation. Enabling smart contracts among different devices and components in the smart grid infrastructure provides optimized grid operations and accelerates global energy transformation with the least transaction costs \cite{mengelkamp2018blockchain}. Blockchain induces consumers to moves toward prosumers (who consumes and produce simultaneously) facilitating them to save electricity cost by selling abundant power to the grid with autonomous safe recording information through smart contracts \cite{andoni2019blockchain}.

\textit{\\Agriculture Field:} 

The agriculture sectors are largely depending on several external factors such as climate, water supply availability, reap quality, complex supply chains, etc. This complexity can be addressed transparently using blockchain technology that could ease the path from the agricultural farm to the supermarket, giving buyers a sense of control and security \cite{yiannas2018new}. In addition, BCT permits small business farms with low assets, can easily access a similar set of data likewise rich farms which making the trading market more transparent. 

\textit{\\Retail and E-commerce:} 

E-commerce based IoT deployment is the emerging business sector which demands the transactions to be autonomous, secure, and lightweight \cite{liu2019anonymous}. The necessity of blockchain in retail industry can be illustrated in several steps of product tracking. A detail of production including time, place, quantity from the manufacturer is listed in the blockchain. Then, information of shipped products is properly logged onto the BC database during transit and data is then updated at the warehouse end with logistic details. Thereafter, wholesaler and retailer dealers update the received product information in the BC. Finally, consumers make anonymous feedback on the products that are recorded in the blockchain. This set of complete procedure brings a good reputation for companies and downsize the overall computational overhead \cite{liu2018normachain}. BCT essentially eliminate some difficulties experienced in the e-commerce and retail business by providing trust, values, and incentive to the consumers. Such business sectors demand more robust research in automated and secured transactions in blockchain platform.

\textit{\\Supply Chains:} 

BCT can considerably reduce time delays and human slip-ups and help to verify the product authentication by tracing their origin. Data accessibility and immutability aspects of blockchain increase the reliability, efficiency, and transparency of the whole supply chain industry \cite{perboli2018blockchain}. Everyday traders, merchants, shippers, and other logistic support personnel handle plenty of choices while shipping a massive number of products across worldwide, each progression of the journey requires extensive paperwork and communication management. In spite of technological advancement, logistics management still needs improvement for keeping and tracking all the records transparently. Blockchain solves these complexity and fragmented tasks by tracking resource, recording exchange, developing effective product management framework for dealing with all archives involved with the logistics procedure \cite{mondal2019blockchain}.

\textit{\\Healthcare:} 

In the healthcare industry, the data for critical patients can be shared effectively using blockchain, which can eliminate errors in patient care and potentially enhance healthcare services \cite{esposito2018blockchain}. Interoperability feature offered by blockchain provides the stakeholders (e.g., healthcare providers, doctors) to access the patient’s health data with proper authentication in a secured manner. Moreover, BC driven IoT system accelerate the diseases management and monitoring facilities, for example, smart pills taking, tracking vital signs, condition investigation remotely, side effects alarm, and providing feedback using smart wearable devices, and quality control \cite{wu2018toward}. The fundamental challenge of maintaining patients’ records is that leaving the health-related information scattered across different healthcare providers. Electronic health records cannot ensure to manage lifelong medical records, accessible across various health service providers. The lack of data management and coordination among medical service providers create additional obstacles for accessing the records. Blockchain helps in providing a secured data sharing structure where any entity can gather all the information of respective patients including their personal details, medicare, medical history, etc.

\subsection{Summary and Lessons Learned}		

The integration of blockchain technology (BCT) into IoT and 6G networks have clearly demonstrated a great potential. The fundamental features of BCT include transparency, immutability, security, anonymity, and the least processing time. BCT does not need to involve a trusted third party or any central intermediary authority in the BC architecture and hence, it minimizes the processing time and cost while performing P2P transactions. Shifting the blockchain paradigm toward a distributed and digitized economy allows smart living with a greater level of integrity in the future. The use of blockchain technology in next generation ultra high-speed 6G networks and massive IoT systems attain the expected level of services and ultra-high performances steering to a positive feedback loop. BCT reduces manmade errors where users’ identity and device are authenticated prior to obtaining access to the network. Besides, each participant has direct control of its own information, and all participated nodes are authenticated, thereby, BC technology enhanced public trust. Blockchain enabled 6G will support a user-friendly environment of intelligent big data analysis for sustainable high-end products and services to address specialized demands by the consumers. A shared BC ledger marketplace can be employed in the asset transactions among multiple peers for network resource sharing, roaming, neutral hosting, network slices, micro-operators, local licensing, software defined networking (SDN), and network function virtualization (NFV) based networking and specialized services. In the case of industry 4.0 and industrial IoT applications, BC empowered sustainable business models will provide extreme reliability, privacy-aware secure communications, zero touch assistance with the advent of green networking, zero carbon technologies. Additionally, blockchain empowered decentralized architectures offers excellent solution in 6G business models in terms of service agility, orchestration, open collaboration, and zero touch management. Moreover, BC based 6G business models offer cloud storage for over the top (OTT) companies, cognitive services such as context aware operations, AI, UAV as a service together with calling and data connectivity features, envisioned to attract customers and increase their profit. Furthermore, blockchain enabled distributed IoT systems address the challenges of conventional IoT architectures by means of advanced efficacy, interoperability, network management, privacy, and security vulnerabilities, particularly in the high degree of complex data handling.

Despite the numerous potential advantages of BC enabled 6G communications and IoT system, the considered architecture is suffering distinct types of shortcomings for implementation. Prior joining to the blockchain, a new block is validated and agreed upon by the active participants in a transparent way. All the participant node requires the witness of prior data for the validation purpose. Moreover, the immutability feature of blockchain restricts the modification and deletion of confirmed transactions. BCT supports distributed applications avoiding single point failure and develop augmentation of trust. As a consequence, blockchain is broadly regarded as a trusted technology in the future generation, but it suffers privacy issues. The appropriate consensus algorithms with asymmetric encryption overcome the challenges of security and privacy aspects. Smart contracts in blockchain allow software defined contracts among users as transactions and hence, the flexible smart contracts introduce multiple distinctive new attacks in the network. However, different cryptographic algorithms such as ring architecture, zero knowledge proof, coin mixing can be used for privacy and security. The reliability of transactions can be achieved using asymmetric encryption mechanisms, hash inclusions and digital signatures. The data communications among multiple operational platforms are paramount concerns in BC enabled cross layer 6G and IoT architectures. Ensuring the least level of latency is another key challenge for BC based high speed system implementation. The communication overhead and computational complexity introduced by blockchain may impose a great challenge for the deployment of billions number of smart devices and corresponding huge transactions experienced by 6G networks and IoT systems.In addition, the energy consumption of IoT devices, least computational power, low storage capacity, lack of policies for data security, and compatibility with 6G are some others research challenges IoT enabled 6G devices. Majority studies related to BC based 6G networks and massive IoT systems are not experimentally demonstrated, provide theoretical analysis with some simulation works. BoT and BC-6G system cover a wide area ranging from individual, industry to worldwide business models. Multiple studies have discussed numerous solutions to resolve distinctive technical issues, however, many aspects remain unexplored.

\section{Conclusions}		\label{sec7}

With the rapid pace of a massive number of IoT devices, a number of obstacles incur to downsize the empowering IoT across various applications. IoT device owners and telecom operators must trust the intermediary node under the centralized architectures. Blockchain technology is an emergent distributed guarantor in terms of a cost-effective solution, nonrepudiation, convergence between embedded technologies, and exploiting traceability to verify the unauthorized action affecting the sustainability charter. This survey paper explored the recent research works directed on the applications of BC in 6G networks and multiple potentials in industrial IoT areas from the perspective of different commercial implementations. We provide an overview of the blockchain features, potential, drivers, requirements for massive IoT enabled by 6G addressing numerous challenges such as integrity of personal data protection, data secrecy, scalability of block contents, adoption cost, privacy of other entities, and governmental regulations. Then a discussion on core technical elements of blockchain including accessing strategies, extensive comparison of consensus algorithms to fit the desired applications are carried out. A sustainable ecosystem focused business model driven by blockchain empowered 6G networks is thoroughly analyzed to address the contemporary global economic crisis. In addition to support in economic performance, a smart contract regard as a core component of BCT that curtail the negative external aspects during manufacturing, enhance the quality of life, promote employability, uplift societal and environmental performances under a wide range of industrial applications. The use of blockchain in IoT industries and 6G use case applications are expected to grow and benefit of smart living and industry sectors mitigating security threats and provide integrity solutions. Blockchain deployment is not without limitations, a lot of challenges lay ahead, and hence, necessary care must be taken to deal with a number of BC challenges to maintain desired QoS, performance level, and adequate security. We listed the research challenges and future directions, outlined numerous possible solutions to address the challenges. It is envisioning that AI, ML, intelligent surfaces, cybertwin, cell-free architecture will likely become promising candidates in future blockchain implementations. This survey paper provides valuable resource for clear understanding the contemporary research contribution of blockchain technology in different implications and is anticipated to persuade further effort for the imminent deployment of next generation wireless networks and industries in the forthcoming years.

\bibliographystyle{IEEEtran}
\bibliography{ref_BC}

\begin{thebibliography}{100}
\providecommand{\url}[1]{#1}
\csname url@samestyle\endcsname
\providecommand{\newblock}{\relax}
\providecommand{\bibinfo}[2]{#2}
\providecommand{\BIBentrySTDinterwordspacing}{\spaceskip=0pt\relax}
\providecommand{\BIBentryALTinterwordstretchfactor}{4}
\providecommand{\BIBentryALTinterwordspacing}{\spaceskip=\fontdimen2\font plus
\BIBentryALTinterwordstretchfactor\fontdimen3\font minus
  \fontdimen4\font\relax}
\providecommand{\BIBforeignlanguage}[2]{{%
\expandafter\ifx\csname l@#1\endcsname\relax
\typeout{** WARNING: IEEEtran.bst: No hyphenation pattern has been}%
\typeout{** loaded for the language `#1'. Using the pattern for}%
\typeout{** the default language instead.}%
\else
\language=\csname l@#1\endcsname
\fi
#2}}
\providecommand{\BIBdecl}{\relax}
\BIBdecl

\bibitem{zhang2017software}
N.~Zhang, P.~Yang, S.~Zhang, D.~Chen, W.~Zhuang, B.~Liang, and X.~S. Shen,
  ``Software defined networking enabled wireless network virtualization:
  {Challenges} and solutions,'' \emph{IEEE Network}, vol.~31, no.~5, pp.
  42--49, 2017.

\bibitem{zhang20196g}
Z.~Zhang, Y.~Xiao, Z.~Ma, M.~Xiao, Z.~Ding, X.~Lei, G.~K. Karagiannidis, and
  P.~Fan, ``{6G} wireless networks: {Vision}, requirements, architecture, and
  key technologies,'' \emph{IEEE Vehicular Technology Magazine}, vol.~14,
  no.~3, pp. 28--41, 2019.

\bibitem{guo2021enabling}
F.~Guo, F.~R. Yu, H.~Zhang, X.~Li, H.~Ji, and V.~C. Leung, ``Enabling massive
  {IoT} toward {6G}: {A} comprehensive survey,'' \emph{IEEE Internet of Things
  Journal}, vol.~8, no.~15, pp. 11\,891--11\,915, 2021.

\bibitem{chowdhury20206g}
M.~Z. Chowdhury, M.~Shahjalal, S.~Ahmed, and Y.~M. Jang, ``{6G} wireless
  communication systems: {Applications}, requirements, technologies,
  challenges, and research directions,'' \emph{IEEE Open Journal of the
  Communications Society}, vol.~1, pp. 957--975, 2020.

\bibitem{tariq2020speculative}
F.~Tariq, M.~R. Khandaker, K.-K. Wong, M.~A. Imran, M.~Bennis, and M.~Debbah,
  ``A speculative study on {6G},'' \emph{IEEE Wireless Communications},
  vol.~27, no.~4, pp. 118--125, 2020.

\bibitem{saad2019vision}
W.~Saad, M.~Bennis, and M.~Chen, ``A vision of {6G} wireless systems:
  {Applications}, trends, technologies, and open research problems,''
  \emph{IEEE network}, vol.~34, no.~3, pp. 134--142, 2020.

\bibitem{qi2020integration}
Q.~Qi, X.~Chen, C.~Zhong, and Z.~Zhang, ``Integration of energy, computation
  and communication in {6G} cellular {Internet of Things},'' \emph{IEEE
  Communications Letters}, vol.~24, no.~6, pp. 1333--1337, 2020.

\bibitem{nguyen2020blockchain}
D.~C. Nguyen, P.~N. Pathirana, M.~Ding, and A.~Seneviratne, ``Blockchain for
  {5G} and beyond networks: {A} state of the art survey,'' \emph{Journal of
  Network and Computer Applications}, vol. 166, p. 102693, 2020.

\bibitem{rawat2019fusion}
D.~B. Rawat, ``Fusion of software defined networking, edge computing, and
  blockchain technology for wireless network virtualization,'' \emph{IEEE
  Communications Magazine}, vol.~57, no.~10, pp. 50--55, 2020.

\bibitem{ahokangas2019business}
P.~Ahokangas, M.~Matinmikko-Blue, S.~Yrj{\"o}l{\"a}, V.~Sepp{\"a}nen,
  H.~H{\"a}mm{\"a}inen, R.~Jurva, and M.~Latva-aho, ``Business models for local
  {5G} micro operators,'' \emph{IEEE Transactions on Cognitive Communications
  and Networking}, vol.~5, no.~3, pp. 730--740, 2019.

\bibitem{dinh2018untangling}
T.~T.~A. Dinh, R.~Liu, M.~Zhang, G.~Chen, B.~C. Ooi, and J.~Wang, ``Untangling
  blockchain: {A} data processing view of blockchain systems,'' \emph{IEEE
  transactions on knowledge and data engineering}, vol.~30, no.~7, pp.
  1366--1385, 2018.

\bibitem{weiss2019application}
M.~B. Weiss, K.~Werbach, D.~C. Sicker, and C.~E.~C. Bastidas, ``On the
  application of blockchains to spectrum management,'' \emph{IEEE Transactions
  on Cognitive Communications and Networking}, vol.~5, no.~2, pp. 193--205,
  2019.

\bibitem{maksymyuk2020blockchain}
T.~Maksymyuk, J.~Gazda, M.~Volosin, G.~Bugar, D.~Horvath, M.~Klymash, and
  M.~Dohler, ``Blockchain-empowered framework for decentralized network
  management in {6G},'' \emph{IEEE Communications Magazine}, vol.~58, no.~9,
  pp. 86--92, 2020.

\bibitem{alladi2019blockchain}
T.~Alladi, V.~Chamola, R.~M. Parizi, and K.-K.~R. Choo, ``Blockchain
  applications for industry 4.0 and industrial {IoT}: {A} review,'' \emph{IEEE
  Access}, vol.~7, pp. 176\,935--176\,951, 2019.

\bibitem{puthal2018everything}
D.~Puthal, N.~Malik, S.~P. Mohanty, E.~Kougianos, and G.~Das, ``Everything you
  wanted to know about the blockchain: {Its} promise, components, processes,
  and problems,'' \emph{IEEE Consumer Electronics Magazine}, vol.~7, no.~4, pp.
  6--14, 2018.

\bibitem{taylor2020systematic}
P.~J. Taylor, T.~Dargahi, A.~Dehghantanha, R.~M. Parizi, and K.-K.~R. Choo, ``A
  systematic literature review of blockchain cyber security,'' \emph{Digital
  Communications and Networks}, vol.~6, no.~2, pp. 147--156, 2020.

\bibitem{efanov2018all}
D.~Efanov and P.~Roschin, ``The all-pervasiveness of the blockchain
  technology,'' \emph{Procedia Computer Science}, vol. 123, pp. 116--121, 2018.

\bibitem{panarello2018blockchain}
A.~Panarello, N.~Tapas, G.~Merlino, F.~Longo, and A.~Puliafito, ``Blockchain
  and {IoT} integration: {A} systematic survey,'' \emph{Sensors}, vol.~18,
  no.~8, p. 2575, 2018.

\bibitem{hang2019design}
L.~Hang and D.-H. Kim, ``Design and implementation of an integrated {IoT}
  blockchain platform for sensing data integrity,'' \emph{Sensors}, vol.~19,
  no.~10, p. 2228, 2019.

\bibitem{singh2020blockiotintelligence}
S.~K. Singh, S.~Rathore, and J.~H. Park, ``Blockiotintelligence: {A}
  blockchain-enabled intelligent {IoT} architecture with artificial
  intelligence,'' \emph{Future Generation Computer Systems}, vol. 110, pp.
  721--743, 2020.

\bibitem{huang2019survey}
T.~Huang, W.~Yang, J.~Wu, J.~Ma, X.~Zhang, and D.~Zhang, ``A survey on green
  {6G} network: {Architecture} and technologies,'' \emph{IEEE Access}, vol.~7,
  pp. 175\,758--175\,768, 2019.

\bibitem{sharma2019toward}
S.~K. Sharma and X.~Wang, ``Toward massive machine type communications in
  ultra-dense cellular {IoT} networks: {Current} issues and machine
  learning-assisted solutions,'' \emph{IEEE Communications Surveys \&
  Tutorials}, vol.~22, no.~1, pp. 426--471, 2020.

\bibitem{hussain2020machine}
F.~Hussain, S.~A. Hassan, R.~Hussain, and E.~Hossain, ``Machine learning for
  resource management in cellular and {IoT} networks: {Potentials}, current
  solutions, and open challenges,'' \emph{IEEE Communications Surveys \&
  Tutorials}, vol.~22, no.~2, pp. 1251--1275, 2020.

\bibitem{fernandez2018review}
T.~M. Fern{\'a}ndez-Caram{\'e}s and P.~Fraga-Lamas, ``A review on the use of
  {Blockchain} for the {Internet of Things},'' \emph{IEEE Access}, vol.~6, pp.
  32\,979--33\,001, 2018.

\bibitem{ferrag2018blockchain}
M.~A. Ferrag, M.~Derdour, M.~Mukherjee, A.~Derhab, L.~Maglaras, and H.~Janicke,
  ``Blockchain technologies for the internet of things: {Research} issues and
  challenges,'' \emph{IEEE Internet of Things Journal}, vol.~6, no.~2, pp.
  2188--2204, 2018.

\bibitem{shen2018blockchain}
C.~Shen and F.~Pena-Mora, ``Blockchain for cities—{A} systematic literature
  review,'' \emph{IEEE Access}, vol.~6, pp. 76\,787--76\,819, 2018.

\bibitem{salman2018security}
T.~Salman, M.~Zolanvari, A.~Erbad, R.~Jain, and M.~Samaka, ``Security services
  using blockchains: {A} state of the art survey,'' \emph{IEEE Communications
  Surveys \& Tutorials}, vol.~21, no.~1, pp. 858--880, 2019.

\bibitem{fraga2019review}
P.~Fraga-Lamas and T.~M. Fern{\'a}ndez-Caram{\'e}s, ``A review on blockchain
  technologies for an advanced and cyber-resilient automotive industry,''
  \emph{IEEE access}, vol.~7, pp. 17\,578--17\,598, 2019.

\bibitem{lu2019blockchain}
H.~Lu, K.~Huang, M.~Azimi, and L.~Guo, ``Blockchain technology in the oil and
  gas industry: {A} review of applications, opportunities, challenges, and
  risks,'' \emph{IEEE Access}, vol.~7, pp. 41\,426--41\,444, 2019.

\bibitem{yang2019integrated}
R.~Yang, F.~R. Yu, P.~Si, Z.~Yang, and Y.~Zhang, ``Integrated blockchain and
  edge computing systems: {A} survey, some research issues and challenges,''
  \emph{IEEE Communications Surveys \& Tutorials}, vol.~21, no.~2, pp.
  1508--1532, 2019.

\bibitem{xie2019survey}
J.~Xie, H.~Tang, T.~Huang, F.~R. Yu, R.~Xie, J.~Liu, and Y.~Liu, ``A survey of
  blockchain technology applied to smart cities: {Research} issues and
  challenges,'' \emph{IEEE Communications Surveys \& Tutorials}, vol.~21,
  no.~3, pp. 2794--2830, 2019.

\bibitem{elisa2018framework}
N.~Elisa, L.~Yang, F.~Chao, and Y.~Cao, ``A framework of blockchain-based
  secure and privacy-preserving e-government system,'' \emph{Wireless
  Networks}, pp. 1--11, 2019.

\bibitem{wei2019convergence}
L.~Wei, J.~Wu, C.~Long, and Y.-B. Lin, ``The convergence of {IoE} and
  blockchain: {Security} challenges,'' \emph{IT Professional}, vol.~21, no.~5,
  pp. 26--32, 2019.

\bibitem{lage2019blockchain}
O.~Lage, ``Blockchain: From {Industry 4.0} to the machine economy,'' in
  \emph{Computer Security Threats}.\hskip 1em plus 0.5em minus 0.4em\relax
  IntechOpen, 2019.

\bibitem{elmamy2020survey}
S.~B. ElMamy, H.~Mrabet, H.~Gharbi, A.~Jemai, and D.~Trentesaux, ``A survey on
  the usage of blockchain technology for cyber-threats in the context of
  {Industry 4.0},'' \emph{Sustainability}, vol.~12, no.~21, p. 9179, 2020.

\bibitem{singh2020blockchain}
M.~Singh, ``Blockchain technology for data management in {Industry 4.0},'' in
  \emph{Blockchain Technology for Industry 4.0}.\hskip 1em plus 0.5em minus
  0.4em\relax Springer, 2020, pp. 59--72.

\bibitem{sekaran2020survival}
R.~Sekaran, R.~Patan, A.~Raveendran, F.~Al-Turjman, M.~Ramachandran, and
  L.~Mostarda, ``Survival study on blockchain based {6G}-enabled mobile edge
  computation for {IoT} automation,'' \emph{IEEE Access}, vol.~8, pp.
  143\,453--143\,463, 2020.

\bibitem{xu2020blockchain}
H.~Xu, P.~V. Klaine, O.~Onireti, B.~Cao, M.~Imran, and L.~Zhang,
  ``Blockchain-enabled resource management and sharing for {6G}
  communications,'' \emph{Digital Communications and Networks}, vol.~6, no.~3,
  pp. 261--269, 2021.

\bibitem{kumari2021amalgamation}
A.~Kumari, R.~Gupta, and S.~Tanwar, ``Amalgamation of blockchain and {IoT} for
  smart cities underlying {6G} communication: {A} comprehensive review,''
  \emph{Computer Communications}, vol. 172, no.~1, pp. 102--118, 2021.

\bibitem{wang2021blockchain}
J.~Wang, X.~Ling, Y.~Le, Y.~Huang, and X.~You, ``Blockchain enabled wireless
  communications: {A} new paradigm towards {6G},'' \emph{National Science
  Review}, 2021.

\bibitem{jiang2021road}
W.~Jiang, B.~Han, M.~A. Habibi, and H.~D. Schotten, ``The road towards {6G}:
  {A} comprehensive survey,'' \emph{IEEE Open Journal of the Communications
  Society}, vol.~2, pp. 334--366, 2021.

\bibitem{cao2020performance}
B.~Cao, Z.~Zhang, D.~Feng, S.~Zhang, L.~Zhang, M.~Peng, and Y.~Li,
  ``Performance analysis and comparison of {PoW, PoS and DAG} based
  blockchains,'' \emph{Digital Communications and Networks}, vol.~6, no.~4, pp.
  480--485, 2020.

\bibitem{xiao2020survey}
Y.~Xiao, N.~Zhang, W.~Lou, and Y.~T. Hou, ``A survey of distributed consensus
  protocols for blockchain networks,'' \emph{IEEE Communications Surveys \&
  Tutorials}, vol.~22, no.~2, pp. 1432--1465, 2020.

\bibitem{yrjola2020could}
S.~Yrj{\"o}l{\"a}, ``How could blockchain transform {6G} towards open
  ecosystemic business models?'' in \emph{IEEE International Conference on
  Communications Workshops (ICC Workshops)}.\hskip 1em plus 0.5em minus
  0.4em\relax IEEE, 2020, pp. 1--6.

\bibitem{nguyen2020privacy}
T.~Nguyen, N.~Tran, L.~Loven, J.~Partala, M.-T. Kechadi, and S.~Pirttikangas,
  ``Privacy-aware blockchain innovation for {6G}: Challenges and
  opportunities,'' in \emph{6G Wireless Summit (6G SUMMIT)}.\hskip 1em plus
  0.5em minus 0.4em\relax IEEE, 2020, pp. 1--5.

\bibitem{ling2019blockchain}
X.~Ling, J.~Wang, T.~Bouchoucha, B.~C. Levy, and Z.~Ding, ``Blockchain radio
  access network {(B-RAN)}: Towards decentralized secure radio access
  paradigm,'' \emph{IEEE Access}, vol.~7, pp. 9714--9723, 2019.

\bibitem{hang2019sla}
L.~Hang and D.-H. Kim, ``{SLA}-based sharing economy service with smart
  contract for resource integrity in the internet of things,'' \emph{Applied
  Sciences}, vol.~9, no.~17, p. 3602, 2019.

\bibitem{xu2018industry}
L.~D. Xu, E.~L. Xu, and L.~Li, ``Industry 4.0: state of the art and future
  trends,'' \emph{International Journal of Production Research}, vol.~56,
  no.~8, pp. 2941--2962, 2018.

\bibitem{dai2019blockchain}
Y.~Dai, D.~Xu, S.~Maharjan, Z.~Chen, Q.~He, and Y.~Zhang, ``Blockchain and deep
  reinforcement learning empowered intelligent {5G} beyond,'' \emph{IEEE
  network}, vol.~33, no.~3, pp. 10--17, 2019.

\bibitem{qiu2019blockchain}
J.~Qiu, D.~Grace, G.~Ding, J.~Yao, and Q.~Wu, ``Blockchain-based secure
  spectrum trading for unmanned-aerial-vehicle-assisted cellular networks: {An}
  operator’s perspective,'' \emph{IEEE Internet of Things Journal}, vol.~7,
  no.~1, pp. 451--466, 2020.

\bibitem{zhou2020blockchain}
Z.~Zhou, X.~Chen, Y.~Zhang, and S.~Mumtaz, ``Blockchain-empowered secure
  spectrum sharing for {5G} heterogeneous networks,'' \emph{IEEE Network},
  vol.~34, no.~1, pp. 24--31, 2020.

\bibitem{bugar2020techno}
G.~Bug{\'a}r, M.~Volo{\v{s}}in, T.~Maksymyuk, J.~Zausinov{\'a}, V.~Gazda,
  D.~Horv{\'a}th, and J.~Gazda, ``Techno-economic framework for dynamic
  operator selection in a multi-tier heterogeneous network,'' \emph{Ad Hoc
  Networks}, vol.~97, p. 102007, 2020.

\bibitem{li2017channel}
A.~Li, G.~Han, J.~J. Rodrigues, and S.~Chan, ``Channel hopping protocols for
  dynamic spectrum management in {5G} technology,'' \emph{IEEE Wireless
  Communications}, vol.~24, no.~5, pp. 102--109, 2018.

\bibitem{liang2020blockchain}
Y.-C. Liang, ``Blockchain for dynamic spectrum management,'' in \emph{Dynamic
  Spectrum Management}.\hskip 1em plus 0.5em minus 0.4em\relax Springer, 2020,
  pp. 121--146.

\bibitem{nour2019blockchain}
B.~Nour, A.~Ksentini, N.~Herbaut, P.~A. Frangoudis, and H.~Moungla, ``A
  blockchain-based network slice broker for {5G} services,'' \emph{IEEE
  Networking Letters}, vol.~1, no.~3, pp. 99--102, 2019.

\bibitem{refaey2019blockchain}
A.~Refaey, K.~Hammad, S.~Magierowski, and E.~Hossain, ``A blockchain policy and
  charging control framework for roaming in cellular networks,'' \emph{IEEE
  Network}, vol.~34, no.~3, pp. 170--177, 2020.

\bibitem{hewa2020role}
T.~Hewa, G.~G{\"u}r, A.~Kalla, M.~Ylianttila, A.~Bracken, and M.~Liyanage,
  ``The role of blockchain in {6G}: challenges, opportunities and research
  directions,'' in \emph{6G Wireless Summit}.\hskip 1em plus 0.5em minus
  0.4em\relax IEEE, 2020, pp. 1--5.

\bibitem{porambage2019sec}
P.~Porambage, T.~Kumar, M.~Liyanage, J.~Partala, L.~Lov{\'e}n, M.~Ylianttila,
  and T.~Sepp{\"a}nen, ``Sec-{EdgeAI}: {AI} for edge security vs security for
  edge {AI},'' \emph{6G Wireless Summit,(Levi, Finland)}, 2019.

\bibitem{bhat20216g}
J.~R. Bhat and S.~A. Alqahtani, ``{6G} ecosystem: current status and future
  perspective,'' \emph{IEEE Access}, vol.~9, pp. 43\,134--43\,167, 2021.

\bibitem{yrjola2020sustainability}
S.~Yrj{\"o}l{\"a}, P.~Ahokangas, and M.~Matinmikko-Blue, ``Sustainability as a
  challenge and driver for novel ecosystemic {6G} business scenarios,''
  \emph{Sustainability}, vol.~12, no.~21, p. 8951, 2020.

\bibitem{imoize20216g}
A.~L. Imoize, O.~Adedeji, N.~Tandiya, and S.~Shetty, ``{6G} enabled smart
  infrastructure for sustainable society: {Opportunities}, challenges, and
  research roadmap,'' \emph{Sensors}, vol.~21, no.~5, p. 1709, 2021.

\bibitem{hassan2019current}
W.~H. Hassan \emph{et~al.}, ``Current research on internet of things {(IoT)}
  security: {A} survey,'' \emph{Computer networks}, vol. 148, pp. 283--294,
  2019.

\bibitem{jesus2018survey}
E.~F. Jesus, V.~R. Chicarino, C.~V. De~Albuquerque, and A.~A. d.~A. Rocha, ``A
  survey of how to use blockchain to secure internet of things and the stalker
  attack,'' \emph{Security and Communication Networks}, vol. 2018, 2018.

\bibitem{qu2018blockchain}
C.~Qu, M.~Tao, J.~Zhang, X.~Hong, and R.~Yuan, ``Blockchain based credibility
  verification method for {IoT} entities,'' \emph{Security and Communication
  Networks}, vol. 2018, 2018.

\bibitem{maroufi2019convergence}
M.~Maroufi, R.~Abdolee, and B.~M. Tazekand, ``On the convergence of blockchain
  and internet of things ({IoT}) technologies,'' \emph{Wireless
  Communications}, vol.~14, no.~1, pp. 1--11, 2019.

\bibitem{lee2018towards}
C.~Lee, L.~Nkenyereye, N.~Sung, and J.~Song, ``Towards a blockchain-enabled
  {IoT} platform using one {M2M} standards,'' in \emph{International Conference
  on Information and Communication Technology Convergence (ICTC)}.\hskip 1em
  plus 0.5em minus 0.4em\relax IEEE, 2019, pp. 97--102.

\bibitem{damianou2019architecture}
A.~Damianou, C.~M. Angelopoulos, and V.~Katos, ``An architecture for blockchain
  over edge-enabled {IoT} for smart circular cities,'' in \emph{International
  Conference on Distributed Computing in Sensor Systems (DCOSS)}.\hskip 1em
  plus 0.5em minus 0.4em\relax IEEE, 2019, pp. 465--472.

\bibitem{mistry2020blockchain}
I.~Mistry, S.~Tanwar, S.~Tyagi, and N.~Kumar, ``Blockchain for {5G}-enabled
  {IoT} for industrial automation: {A} systematic review, solutions, and
  challenges,'' \emph{Mechanical Systems and Signal Processing}, vol. 135, p.
  106382, 2020.

\bibitem{choi2019multiple}
B.-G. Choi, E.~Jeong, and S.-W. Kim, ``Multiple security certification system
  between blockchain based terminal and internet of things device:
  {Implication} for open innovation,'' \emph{Journal of Open Innovation:
  Technology, Market, and Complexity}, vol.~5, no.~4, p.~87, 2020.

\bibitem{biswal2020authenticating}
A.~K. Biswal, P.~Maiti, S.~Bebarta, B.~Sahoo, and A.~K. Turuk, ``Authenticating
  {IoT} devices with blockchain,'' in \emph{Advanced Applications of Blockchain
  Technology}.\hskip 1em plus 0.5em minus 0.4em\relax Springer, 2020, pp.
  177--205.

\bibitem{yaqoob2020blockchain}
I.~Yaqoob, K.~Salah, M.~Uddin, R.~Jayaraman, M.~Omar, and M.~Imran,
  ``Blockchain for digital twins: Recent advances and future research
  challenges,'' \emph{IEEE Network}, vol.~34, no.~5, pp. 290--298, 2020.

\bibitem{liu2020blockchain}
Y.~Liu, F.~R. Yu, X.~Li, H.~Ji, and V.~C. Leung, ``Blockchain and machine
  learning for communications and networking systems,'' \emph{IEEE
  Communications Surveys \& Tutorials}, vol.~22, no.~2, pp. 1392--1431, 2020.

\bibitem{mendki2020blockchain}
P.~Mendki, ``Blockchain enabled iot edge computing: Addressing privacy,
  security and other challenges,'' in \emph{International Conference on
  Blockchain Technology}, 2020, pp. 63--67.

\bibitem{ishmaev2020ethical}
G.~Ishmaev, ``The ethical limits of blockchain-enabled markets for private
  {IoT} data,'' \emph{Philosophy \& Technology}, vol.~33, no.~3, pp. 411--432,
  2020.

\bibitem{franke2020survey}
U.~Franke and J.~Wernberg, ``A survey of cyber security in the {Swedish}
  manufacturing industry,'' in \emph{2020 International Conference on Cyber
  Situational Awareness, Data Analytics and Assessment (CyberSA)}.\hskip 1em
  plus 0.5em minus 0.4em\relax IEEE, 2020, pp. 1--8.

\bibitem{yan2017industrial}
J.~Yan, Y.~Meng, L.~Lu, and L.~Li, ``Industrial big data in an industry 4.0
  environment: {Challenges}, schemes, and applications for predictive
  maintenance,'' \emph{IEEE Access}, vol.~6, pp. 23\,484--23\,491, 2018.

\bibitem{xu2018survey}
H.~Xu, W.~Yu, D.~Griffith, and N.~Golmie, ``A survey on industrial {Internet of
  Things}: {A} cyber-physical systems perspective,'' \emph{IEEE Access},
  vol.~7, pp. 78\,238--78\,259, 2019.

\bibitem{aitzhan2016security}
N.~Z. Aitzhan and D.~Svetinovic, ``Security and privacy in decentralized energy
  trading through multi-signatures, blockchain and anonymous messaging
  streams,'' \emph{IEEE Transactions on Dependable and Secure Computing},
  vol.~15, no.~5, pp. 840--852, 2017.

\bibitem{lin2018id}
Q.~Lin, H.~Yan, Z.~Huang, W.~Chen, J.~Shen, and Y.~Tang, ``An {ID}-based
  linearly homomorphic signature scheme and its application in blockchain,''
  \emph{IEEE Access}, vol.~6, pp. 20\,632--20\,640, 2018.

\bibitem{hammi2018bubbles}
M.~T. Hammi, B.~Hammi, P.~Bellot, and A.~Serhrouchni, ``Bubbles of trust: A
  decentralized blockchain-based authentication system for {IoT},''
  \emph{Computers \& Security}, vol.~78, pp. 126--142, 2018.

\bibitem{lu2018privacy}
Z.~Lu, W.~Liu, Q.~Wang, G.~Qu, and Z.~Liu, ``A privacy-preserving trust model
  based on blockchain for {VANETs},'' \emph{IEEE Access}, vol.~6, pp.
  45\,655--45\,664, 2018.

\bibitem{li2020survey}
X.~Li, P.~Jiang, T.~Chen, X.~Luo, and Q.~Wen, ``A survey on the security of
  blockchain systems,'' \emph{Future Generation Computer Systems}, vol. 107,
  pp. 841--853, 2020.

\bibitem{tang2019future}
F.~Tang, Y.~Kawamoto, N.~Kato, and J.~Liu, ``Future intelligent and secure
  vehicular network toward {6G}: {Machine}-learning approaches,''
  \emph{Proceedings of the IEEE}, vol. 108, no.~2, pp. 292--307, 2020.

\bibitem{mendling2018blockchains}
J.~Mendling, I.~Weber, W.~V.~D. Aalst, J.~V. Brocke, C.~Cabanillas, F.~Daniel,
  S.~Debois, C.~D. Ciccio, M.~Dumas, S.~Dustdar \emph{et~al.}, ``Blockchains
  for business process management-challenges and opportunities,'' \emph{ACM
  Transactions on Management Information Systems (TMIS)}, vol.~9, no.~1, pp.
  1--16, 2018.

\bibitem{dorri2019mof}
A.~Dorri, S.~S. Kanhere, and R.~Jurdak, ``{MOF-BC}: A memory optimized and
  flexible blockchain for large scale networks,'' \emph{Future Generation
  Computer Systems}, vol.~92, pp. 357--373, 2019.

\bibitem{moustafa2019new}
N.~Moustafa, ``New generations of internet of things datasets for cybersecurity
  applications based machine learning: {TON\_IoT} datasets,'' in
  \emph{Proceedings of the eResearch Australasia Conference, Brisbane,
  Australia}, 2019, pp. 21--25.

\bibitem{2dai2019blockchain2}
H.-N. Dai, Z.~Zheng, and Y.~Zhang, ``Blockchain for internet of things: {A}
  survey,'' \emph{IEEE Internet of Things Journal}, vol.~6, no.~5, pp.
  8076--8094, 2019.

\bibitem{mishra2018detailed}
P.~Mishra, V.~Varadharajan, U.~Tupakula, and E.~S. Pilli, ``A detailed
  investigation and analysis of using machine learning techniques for intrusion
  detection,'' \emph{IEEE Communications Surveys \& Tutorials}, vol.~21, no.~1,
  pp. 686--728, 2018.

\bibitem{salah2019blockchain}
K.~Salah, M.~H.~U. Rehman, N.~Nizamuddin, and A.~Al-Fuqaha, ``Blockchain for
  {AI}: {Review} and open research challenges,'' \emph{IEEE Access}, vol.~7,
  pp. 10\,127--10\,149, 2019.

\bibitem{qiu2020ai}
C.~Qiu, H.~Yao, X.~Wang, N.~Zhang, F.~R. Yu, and D.~Niyato, ``{AI}-chain:
  {Blockchain} energized edge intelligence for beyond {5G} networks,''
  \emph{IEEE Network}, vol.~34, no.~6, pp. 62--69, 2020.

\bibitem{jaschke2018unsupervised}
A.~J{\"a}schke and F.~Armknecht, ``Unsupervised machine learning on encrypted
  data,'' in \emph{International Conference on Selected Areas in
  Cryptography}.\hskip 1em plus 0.5em minus 0.4em\relax Springer, 2019, pp.
  453--478.

\bibitem{talapur2018reliable}
G.~G. Talapur, H.~M. Suryawanshi, L.~Xu, and A.~B. Shitole, ``A reliable
  microgrid with seamless transition between grid connected and islanded mode
  for residential community with enhanced power quality,'' \emph{IEEE
  Transactions on Industry Applications}, vol.~54, no.~5, pp. 5246--5255, 2018.

\bibitem{na2020uav}
Z.~Na, Y.~Liu, J.~Shi, C.~Liu, and Z.~Gao, ``{UAV}-supported clustered {NOMA}
  for {6G}-enabled {Internet of Things}: {Trajectory} planning and resource
  allocation,'' \emph{IEEE Internet of Things Journal}, 2020.

\bibitem{li2018uav}
B.~Li, Z.~Fei, and Y.~Zhang, ``{UAV} communications for {5G} and beyond:
  {Recent} advances and future trends,'' \emph{IEEE Internet of Things
  Journal}, vol.~6, no.~2, pp. 2241--2263, 2019.

\bibitem{mengelkamp2018blockchain}
E.~Mengelkamp, B.~Notheisen, C.~Beer, D.~Dauer, and C.~Weinhardt, ``A
  blockchain-based smart grid: {Towards} sustainable local energy markets,''
  \emph{Computer Science-Research and Development}, vol.~33, no.~1, pp.
  207--214, 2018.

\bibitem{andoni2019blockchain}
M.~Andoni, V.~Robu, D.~Flynn, S.~Abram, D.~Geach, D.~Jenkins, P.~McCallum, and
  A.~Peacock, ``Blockchain technology in the energy sector: {A} systematic
  review of challenges and opportunities,'' \emph{Renewable and Sustainable
  Energy Reviews}, vol. 100, pp. 143--174, 2019.

\bibitem{yiannas2018new}
F.~Yiannas, ``A new era of food transparency powered by blockchain,''
  \emph{Innovations: Technology, Governance, Globalization}, vol.~12, no. 1-2,
  pp. 46--56, 2018.

\bibitem{liu2019anonymous}
D.~Liu, A.~Alahmadi, J.~Ni, X.~Lin, and X.~Shen, ``Anonymous reputation system
  for {IIoT}-enabled retail marketing atop {PoS} blockchain,'' \emph{IEEE
  Transactions on Industrial Informatics}, vol.~15, no.~6, pp. 3527--3537,
  2019.

\bibitem{liu2018normachain}
C.~Liu, Y.~Xiao, V.~Javangula, Q.~Hu, S.~Wang, and X.~Cheng, ``Normachain: {A}
  blockchain-based normalized autonomous transaction settlement system for
  {IoT}-based e-commerce,'' \emph{IEEE Internet of Things Journal}, vol.~6,
  no.~3, pp. 4680--4693, 2019.

\bibitem{perboli2018blockchain}
G.~Perboli, S.~Musso, and M.~Rosano, ``Blockchain in logistics and supply
  chain: {A} lean approach for designing real-world use cases,'' \emph{Ieee
  Access}, vol.~6, pp. 62\,018--62\,028, 2018.

\bibitem{mondal2019blockchain}
S.~Mondal, K.~P. Wijewardena, S.~Karuppuswami, N.~Kriti, D.~Kumar, and
  P.~Chahal, ``Blockchain inspired {RFID}-based information architecture for
  food supply chain,'' \emph{IEEE Internet of Things Journal}, vol.~6, no.~3,
  pp. 5803--5813, 2019.

\bibitem{esposito2018blockchain}
C.~Esposito, A.~De~Santis, G.~Tortora, H.~Chang, and K.-K.~R. Choo,
  ``Blockchain: {A} panacea for healthcare cloud-based data security and
  privacy?'' \emph{IEEE Cloud Computing}, vol.~5, no.~1, pp. 31--37, 2018.

\bibitem{wu2018toward}
H.-T. Wu and C.-W. Tsai, ``Toward blockchains for health-care systems:
  {Applying} the bilinear pairing technology to ensure privacy protection and
  accuracy in data sharing,'' \emph{IEEE Consumer Electronics Magazine},
  vol.~7, no.~4, pp. 65--71, 2018.

\end{thebibliography}
\nocite{*}

\end{document}